\documentclass[twocolumn,preprintnumbers,amsmath,amssymb]{revtex4}

\usepackage{dcolumn}
\usepackage{bm}
\usepackage{epsfig,graphicx,psfrag}

\begin{document}

\newcommand{\be}{\begin{equation}}\newcommand{\ee}{\end{equation}}
\newcommand{\bear}{\begin{eqnarray}}\newcommand{\eear}{\end{eqnarray}} 
\newcommand{\ba}{\begin{array}}\newcommand{\ea}{\end{array}}
\newcommand{\lae}{\begin{array}{c}\,\sim\vspace{-1.7em}\\< \end{array}}
\newcommand{\gae}{\begin{array}{c}\,\sim\vspace{-1.58em}\\> \end{array}}
\renewcommand{\theequation}{\arabic{section}.\arabic{equation}}
\newcommand{\lsim}{\begin{array}{c} \sim \vspace{-16pt}\\< \end{array}}

\newcommand{\ZZ}{\ensuremath{\mathcal{\delta{\cal{Z}}}}}


\preprint{\small FERMILAB-Pub-06-009-T, \ \  CU-TP-1143 \ }
\preprint{\small  hep-ph/0601186 \rule{0em}{1.3em} }  

\title{\Large Resonances from Two Universal Extra Dimensions \rule{0em}{2.1em}}

\author{\large Gustavo Burdman$^1$, Bogdan A.~Dobrescu$^2$, Eduardo 
Pont\'{o}n$^3$ \rule{0em}{0.9em}} 

\affiliation{\rule{0em}{0.9em}\\
$^1$Instituto de F\'{i}sica, Universidade de S\~{a}o 
Paulo, 
S\~{a}o Paulo, SP 05508-900, Brazil \\ \rule{0em}{1.6em}
$^2$Fermilab, Batavia, IL 60510, USA \ \ \\ \rule{0em}{1.6em}
$^3$Department of Physics, Columbia University,\\
New York, NY 10027, USA }

\date{January 23, 2006; \ revised: August 20, 2007 \rule{0em}{1.8em}} 

  \begin{abstract}\rule{0em}{1.8em}

Standard model gauge bosons propagating in two universal extra
dimensions give rise to heavy spin-1 and spin-0 particles.
The lightest of these, carrying Kaluza-Klein numbers (1,0), may be 
produced only in pairs at colliders, whereas the (1,1) modes, which 
are heavier by a factor of $\sqrt{2}$, may be singly produced. 
We show that the cascade decays of (1,1) particles generate a series 
of closely-spaced narrow resonances in the $t \bar{t}$ invariant 
mass distribution. At the Tevatron, $s$-channel production of 
(1,1) gluons and electroweak bosons will be sensitive to 
$t \bar{t}$ resonances up to masses in the 0.5--0.7 TeV range.
Searches at the LHC for 
resonances originating from several higher-level modes will 
further test the existence of two universal extra dimensions.
\end{abstract}
\maketitle


\section{Introduction}

If the standard model gauge bosons propagate in extra dimensions, then
for each of the $SU(3)_C \times SU(2)_W \times U(1)_Y$ gauge fields
there is a tower of heavy vector bosons that could produce signals in
collider experiments.  These heavy vector bosons are commonly called
Kaluza-Klein (KK) modes of the gauge bosons \cite{Dienes:1998vg}, and
we will refer to them in what follows as ``vector modes''.  The
properties of the vector modes depend crucially on the number,
compactification and metric of extra dimensions, as well as on what
other fields propagate in the extra dimensions.

For example, if the metric is flat and no quarks or leptons propagate
in the extra dimensions, then vector-mode exchange among fermions
produces too large corrections to the electroweak observables, unless
the compactification scale (which approximately sets the mass of the
lightest vector modes within each tower, called level-1 states) is
higher than roughly 6 TeV \cite{Cheung:2001mq}, pushing the vector
modes beyond the reach of the Large Hadron Collider (LHC)
\cite{Accomando:1999sj}.  By contrast, if the extra dimensions are
universal, {\it i.e.}, all standard model particles propagate in the
extra dimensions, then the limit on the compactification scale is
close to the electroweak scale \cite{Appelquist:2000nn}, so that the 
vector modes could be produced not only at the LHC but also at the Tevatron.  
In that case, however, a KK parity is conserved implying that level-1
vector modes may be produced only in pairs, and that their decays
involve soft leptons and jets plus missing energy \cite{Cheng:2002ab},
making their discovery challenging.

The vector modes that are particularly interesting for collider searches 
are level-2 states from universal extra dimensions.  These
have individual couplings to the observed fermions, induced by loops
within the higher-dimensional effective theory \cite{Cheng:2002iz}, or
by boundary operators generated at the cut-off scale, $M_s$, where
some new physics should smooth out the ultraviolet behavior of the
theory \cite{Appelquist:2000nn}.  The induced couplings are rather
small, being suppressed by either a loop factor or a volume factor, so
that one need not worry about the constraints from electroweak
precision measurements.  At the same time, the suppression may be not
too strong, allowing a potentially sizable $s$-channel production
at high-energy colliders.  This possibility in the case of one 
universal extra dimension, where the level-2 masses are roughly twice 
as large as the level-1 masses, has been noted in
Ref.~\cite{Cheng:2002ab} and analyzed in detail in Ref.~\cite{Datta:2005zs}.

In this paper we point out that level-2 vector modes in the case of
{\it two} universal extra dimensions offer better opportunities for
discovery.  The reason is that the level-2 vector modes in this case
have masses which are larger than the level-1 masses by a factor of
approximately $ \sqrt{2}$.  As a result, their production is possible
at smaller center-of-mass energies, and the decays of level-2 states
into pairs of level-1 states, which would lead to only soft leptons
and jets in the detector, are kinematically forbidden (as opposed to
the case of one universal extra dimension where such decays are
typically allowed).  Then the level-2 states, 
characterized by KK numbers (1,1), have large branching fractions 
for decays into a pair of standard model particles giving rise to a
high $p_T$ signal.  Another distinctive feature of two universal extra
dimensions is that each vector mode is accompanied by a spin-0
particle in the adjoint representation of the corresponding gauge
group.

In Section \ref{sec:6DSM} we present the standard model in two 
universal extra dimensions compactified on the chiral square,
which is the simplest compactification consistent with the 
chirality of the quarks and leptons. We concentrate especially on 
the mass spectrum and KK-number violating interactions of the KK modes.  
In Section \ref{sec:decays} we compute in detail the branching fractions
of the (1,1) modes, which is 
useful for any future phenomenological study of the standard model 
in six dimensions.
We then turn to resonant production of the (1,1) modes at the Tevatron,
and estimate the expected reach of Run II in Section~\ref{sec:tevatron}. 
The more complex phenomenology at the LHC is briefly discussed in
Section~\ref{sec:LHC}, and then our results are summarized in
Section~\ref{sec:conclusions}. 

\section{Six-Dimensional Standard Model}\label{sec:6DSM}
\setcounter{equation}{0}

We consider the standard model in six dimensions, with two dimensions
compactified.  Each of the gauge fields has six components: for
example, the six-dimensional (6D) 
gluon field $G_\alpha^a$, where $a$ labels the eight
$SU(3)_C$ generators, has a 6D Lorentz index $\alpha = 0,1, \ldots, 5$.
The quark and lepton fields are chiral 6D fermions, which have four
components.  The requirements of 6D anomaly cancellations and fermion
mass generation \cite{Dobrescu:2001ae} force the weak-doublet quarks
to have opposite 6D chirality than the weak-singlet quarks, so that
the quarks of one generation are given by $Q_+=(U_+,D_+)$, $U_-$,
$D_-$, where $\pm$ labels the 6D chiralities.  The chirality
assignment in the 6D lepton sector is similar (its implications for
the neutrino masses are analyzed in \cite{Appelquist:2002ft}).

The zero-mode states, which are particles of zero momentum along the
extra dimensions, are identified with the observed standard model
particles.  Since the observed quarks and leptons have definite 4D
chirality, an immediate constraint on any 6D extension of the standard
model is that the compactification of the two extra dimensions must allow
chiral zero-mode fermions.  A simple compactification of this type has
been studied in detail in \cite{Dobrescu:2004zi, Burdman:2005sr}.  It
consists of a square, $0 \leq x_4, x_5 \leq \pi R$, where $x_4,x_5$
are the coordinates of the extra dimensions and $R$ is the
compactification ``radius''.  The compactification is obtained by
imposing the identification of two pairs of adjacent sides of the
square, and we refer to it as the ``chiral square''.

\subsection{KK decomposition}
\label{sec:KKdecomposition}

For any 6D field $\Phi(x^\mu, x^4, x^5)$ that has a zero mode, the
field equations have the following solution:
\be
\Phi = 
\sum_{j,k}   
\left( \cos\frac{j x^4 + k x^5}{R} + \cos\frac{k x^4 - j x^5}{R}
\right) \frac{\Phi^{(j,k)} (x^\mu)}{\pi R(1+\delta_{j,0})}  ~.
\label{fourier}
\ee
The KK numbers, $j$ and $k$, are integers with $j \geq 1$ and $k \geq
0$, or $j=k=0$.  The 4D fields $\Phi^{(j,k)}(x^\mu)$ are the KK modes
of the 6D field $\Phi$.  They have masses due to the momentum along
$x^4, x^5$ given by
\be
\label{kk-mass}
M_{j,k} = \frac{1}{R} \sqrt{j^2 + k^2} ~,
\ee
so that the mass spectrum, in the limit where other contributions to
physical masses are neglected, starts with $M_{0,0} = 0$, $M_{1,0} =
1/R$, $M_{1,1} = \sqrt{2}/R$, $M_{2,0} = 2/R, \ldots$ For 6D fields
that do not have a zero mode, the KK decompositions differ from
Eq.~(\ref{fourier}), as shown in \cite{Dobrescu:2004zi,
Burdman:2005sr}, but their KK mass spectrum is the same for the
massive states.

The 6D gluon and electroweak gauge bosons decompose each into a tower
of 4D spin-1 fields, a tower of 4D spin-0 fields which are eaten by
the heavy spin-1 fields, and a tower of 4D spin-0 fields which remain
in the spectrum, all belonging to the adjoint representation of the
corresponding gauge group.  We refer to these latter spin-0 fields as
``spinless adjoints''.  The zero modes of the spin-1 fields are the
standard model gauge bosons, while the spin-0 fields do not have zero
modes.  Therefore, in the unitary gauge the 6D gluon field includes at
each nonzero KK level a vector mode $G_\mu^{(j,k)a}$ and a real scalar
field $G_H^{(j,k)a}$.  The 6D weak gauge fields have KK modes
$W^{(j,k)\pm}_\mu$, $W^{(j,k)\pm}_H$, $W^{(j,k)3}_\mu$ and
$W^{(j,k)3}_H$, while the hypercharge KK gauge bosons are
$B^{(j,k)}_\mu$, $B^{(j,k)}_H$.  Electroweak symmetry breaking due to
the 6D VEV of the Higgs doublet (as discussed in general in
\cite{Burdman:2005sr}), mixes $W^{(j,k)3}_\mu$ and $B^{(j,k)}_\mu$, as
well as $W^{(j,k)3}_H$ and $B^{(j,k)}_H$.
However, for $1/R \gae 300$ GeV, this mixing is small
\cite{Cheng:2002iz}, and we will neglect it in what follows.  The 6D
Higgs doublet decomposes into a tower of 4D weak doublets.  The
zero-mode doublet gives the longitudinal degrees of freedom of the $W$
and $Z$ and a Higgs boson, while at each nonzero KK level three of the
degrees of freedom of the massive Higgs doublet mix with the
longitudinal components of the electroweak vector modes (this mixing
is also suppressed by $M_Z R$).

The 6D quark and lepton fields decompose each into a tower of heavy
vector-like 4D fermions and a chiral zero mode identified with the
observed fermion.  Explicitly, the standard model quark doublets are
given by $(u_L, d_L) \equiv Q_{+_L}^{(0,0)}$, while the standard
weak-singlet quarks are $u_R \equiv U_{-_R}^{(0,0)}$ and $d_R \equiv
D_{-_R}^{(0,0)}$, where a generation index is implicit.

\subsection{Localized Operators}
\label{sec:local}

The ``chiral square'' compactification is a two-dimensional space
having the topology of a sphere.  It is flat everywhere, with the
exception of conical singularities at the corners of the square.
Altogether there are three such conical singularities, given that the
$(0, \pi R) $ and $(\pi R, 0) $ corners are identified.

Operators localized at the singular points are generated by loops
involving the bulk interactions \cite{Ponton:2005kx}, as in the theories
studied in Ref.~\cite{Georgi:2000ks, Cheng:2002iz}.  The space around
the conical singularities is symmetric under rotations in the
$(x^{4},x^{5})$ plane, and therefore the localized operators have an
$SO(2)$ symmetry.  Furthermore, the bulk interactions are symmetric
under reflections with respect to the center of the square.  This
symmetry is a KK parity, labeled by $Z_2^{\rm KK}$.  It implies that
the operators generated at $(0, 0) $ are identical with those at $(\pi
R, \pi R) $.

Other contributions to the localized operators might be induced by
physics above the cut-off scale.  They should also be $SO(2)$
symmetric.  In addition, it is compelling to assume that these
UV-generated localized operators are $Z_2^{\rm KK}$ symmetric, so
that the stability of the lightest KK particle, a promising dark
matter candidate \cite{Servant:2002aq},    
is ensured.

The 4D Lagrangian can be written as
\bear
\int_0^{\pi R} dx^4 \int_0^{\pi R} dx^5 \left\{ {\cal L}_{\rm bulk} +
\delta (x_4) \delta (\pi R - x_5) {\cal L}_2 \right.
\nonumber \\ \left.
\mbox{} + \left[ \delta (x_4) \delta (x_5) + \delta (\pi R - x_4) 
\delta (\pi R - x_5) \right]
{\cal L}_1 \right\} ~.
\label{l4d}
\eear
${\cal L}_{\rm bulk}$ includes 6D kinetic terms for the quarks,
leptons, $SU(3)_C \times SU(2)_W \times U(1)_Y$ gauge fields, and a
Higgs doublet, 6D Yukawa couplings of the quarks and leptons to the
Higgs doublet, and a 6D Higgs potential.  The form of these terms
can be derived from the general 6D Lagrangians discussed in
\cite{Dobrescu:2004zi, Burdman:2005sr}.  ${\cal L}_1$ and ${\cal L}_2$
contain all the localized operators.  In particular, these include
4D-like kinetic terms for all 6D fields, and the pieces of 6D kinetic
terms that describe motion along the extra dimensions.  For example,
the localized operators of the lowest mass dimension that involve the
6D quark field $U_-$ appear in ${\cal L}_p$ ($p = 1,2$) as
%
\be
\frac{C_{pU}}{2^{2}M_s^2} i
\overline{U}_{-_R} \Gamma^\mu D_\mu U_{-_R} + 
\left(\frac{C_{pU}^\prime}{2^{2}M_s^2}
i \overline{U}_{-_R} \Gamma^l D_l U_{-_L} + {\rm H.c.} \right)~,
\label{bterms}
\ee
where $\Gamma^\mu$ with $\mu = 0,1,2,3$ and $\Gamma^l$ with $l=4,5$
are anti-commuting $8\times8$ matrices, $D_\mu, D_l$ are covariant
derivatives, $C_{pU}$ ($C_{pU}^\prime$) are real (complex)
dimensionless parameters, and $M_s$ is the cut-off scale.  For
convenience, we also wrote explicit factors of $\left(1/2\right)^{2}$
to account for an enhancement due to the values of the wavefunctions
in Eq.~(\ref{fourier}) at the singular points.  The localized
operators of the lowest mass dimension that involve the 6D gluon field
are given by
\be
{\cal L}_{p} \supset - \frac{1}{4} \, \frac{C_{pG}}{2^{2}M_s^2} 
G^{\mu\nu}G_{\mu\nu} 
- \frac{1}{2} \, \frac{C^\prime_{pG}}{2^{2}M_s^2}\left(G_{45}\right)^2 
~,
\label{gluon-terms}
\ee
where $C_{pG}$ and $C^\prime_{pG}$ with $p = 1,2$ are real
dimensionless parameters.

As mentioned above, the contributions
to the localized operators in Eq.~(\ref{bterms}) and
(\ref{gluon-terms}) arise from two sources: loops with KK modes,
and physics above the cut-off scale.
The bare contribution, from physics at or above the cut-off scale, to
the coefficients of the localized terms can be estimated by assuming
that the localized couplings get strong at the cut-off scale $M_s$,
where $M_s$ is the scale at which the QCD interactions
become strong in the ultraviolet.  Using naive dimensional analysis
(NDA) in the 6D theory \cite{Chacko:1999hg}, we estimate the
coefficients $C_{pG}$ and $C_{pG}^\prime$ in Eq.~(\ref{gluon-terms}),
$C_{pU}, {\rm Re} C_{pU}^\prime, {\rm Im} C_{pU}^\prime$ in
Eq.~(\ref{bterms}), and the analogous coefficients associated with the
$Q_+$ and $D_-$ fields, to be all of the order of $l_6/l_4 = 8\pi$,
where $l_{6} = 128 \pi^{3}$ and $l_4 = 16 \pi^{2}$ are 6D and 4D loop
factors, respectively.
This estimate assumes that the localized term receives contributions
from color interactions.  If this is not the case, then there is an
associated suppression.

Furthermore, NDA gives $(\pi R M_s)^{2} \sim l_{6}/(g_s^2 N_{c})$,
where $g_s$ is the 4D QCD gauge coupling and $N_{c} = 3$ is the number
of colors.  Thus, the bare contribution to the effective 4D coupling
is of the order of $(l_{6}/l_{4})/(\pi R M_{s})^{2} \sim
g^2_{s}N_{c}/l_{4}$.  Also, the separation between the
compactification scale and the cut-off scale is
\be
\label{separation}
M_s R \sim \left(\frac{32}{\alpha_s N_c}\right)^{\! 1/2} \approx 10 ~.
\ee
The strong coupling constant is evaluated here at the compactification
scale, $1/R$: $\alpha_s \equiv g_s^2/(4\pi) \approx 0.1$.

The localized operators are also renormalized by the physics below the
cutoff scale, $M_{s}$.  These contributions were calculated in
\cite{Ponton:2005kx} at one-loop order.  For the fermion kinetic terms in
Eq.~(\ref{bterms}) involving 4D derivatives, one obtains
\bear
\frac{C_{1f}}{\left(\pi M_s R\right)^2} = \left[ -4 \sum_{A} g^{2}_{A} C_{2}(f) 
+ \frac{5}{8} \sum_{i} \lambda^{2}_{i} \right]
\frac{1}{16\pi^2} \ln \frac{M_{s}^2}{\mu^2}~,&&
\nonumber \\ [0.7em]
\frac{C_{2f}}{\left(\pi M_s R\right)^2} = \left[ - 2 \sum_{A} g^{2}_{A} C_{2}(f) 
+ \frac{1}{4} \sum_{i} \lambda^{2}_{i} \right]
\frac{1}{16\pi^2} \ln \frac{M_{s}^2}{\mu^2}~,&&
\nonumber \\ [-3em] \nonumber 
\eear
\be
\label{rf}
\ee
where $\lambda_i$ are Yukawa couplings of the fermion $f$ to
{\textit{complex}} scalars having zero modes (the $i$ sum is over the
scalars), $g_A$ is the 4D gauge coupling,  $C_2(A)$ and
$C_2(f)$ are the quadratic Casimir eigenvalues of the gauge fields and
fermions, respectively [for an $SU(N)$ gauge group, $C_2(A) = N$, and
if $f$ is in the fundamental representation, $C_2(f) = (N^2 -
1)/(2N)$], and $T(f)$ and $T(s)$ are the indices of the
representations to which the fermion $f$ and scalar $s$ belong [$T(f)
= T(s) = 1/2$ in the fundamental representation].  Notice that these
contributions are scale dependent, and $\mu$ should be taken of the
order of the characteristic scale of the process of interest.

For the coefficients of the fermion kinetic terms with derivatives in
the plane of the extra dimensions, one finds that only the Yukawa
couplings contribute:
\bear
\frac{C^\prime_{1f}}{\left(\pi M_s R\right)^2} &=& 
\frac{5}{8} \sum_{i} \lambda^{2}_{i}
 \frac{1}{16\pi^2} \ln \frac{M_{s}^2}{\mu^2}~,
\nonumber \\ [0.4em]
\frac{C^\prime_{2f}}{\left(\pi M_s R\right)^2} &=& 
\frac{1}{4} \sum_{i} \lambda^{2}_{i}
\frac{1}{16\pi^2} \ln \frac{M_{s}^2}{\mu^2}~,
\label{rfprime}
\eear
where again the sum runs over complex scalars.  

The
coefficients of the 4D gauge kinetic terms in Eq.~(\ref{gluon-terms})
are found to be
\bear
\frac{C_{1A}}{\left(\pi M_s R\right)^2} &=& \left[ -\frac{14}{3} C_{2}(A) + 
\frac{2}{3} \sum_{f} T(f) + \frac{5}{12} \sum_{s} T(s) \right]
\nonumber \\
& & \times  \frac{g^{2}_{A}}{16\pi^2} 
\ln \frac{M_{s}^2}{\mu^2}~,
\nonumber \\ [0.4em]
\frac{C_{2A}}{\left(\pi M_s R\right)^2} &=& \left[ - 2\, C_{2}(A) + \frac{1}{6} \sum_{s} T(s) \right]
\frac{g^{2}_{A}}{16\pi^2} \ln \frac{M_{s}^2}{\mu^2}~,
\nonumber \\ 
\label{rAg}
\eear
where $A$ stands for any gauge field.  The sum over $f$ involves all
6D Weyl fermions having a zero mode of any 4D chirality, while the 
sum over $s$ involves all complex scalars having a zero mode.  

Finally, for the operators involving
the spinless adjoints $A_{4}$--$A_{5}$ one finds
\bear
\frac{C'_{1A}}{\left(\pi M_s R\right)^2} &=& \left[ 8 C_{2}(A) - 4 \sum_{f} T(f) 
+ \frac{13}{4} \sum_{s} T(s) \right]
\nonumber \\
& & \times \frac{g^{2}_{A}}{16\pi^2} \ln \frac{M_{s}^2}{\mu^2}~,
\nonumber \\ [0.4em]
\frac{C'_{2A}}{\left(\pi M_s R\right)^2} &=& \left[ 2\, C_{2}(A) + 
\frac{1}{2} \sum_{s} T(s) \right]
\frac{g^{2}_{A}}{16\pi^2} \ln \frac{M_{s}^2}{\mu^2}~,
\nonumber \\
\label{rA}
\eear
where again the sums run over 6D Weyl fermions and complex scalars
having zero modes.

We note that the contribution due to the physics below the cutoff
scale is enhanced by a logarithmic factor compared to the ``bare''
contributions from physics at or above $M_{s}$.  However, in the
present class of models the logarithm is at most a few [see
Eq.~(\ref{separation})].  Also, for strongly interacting particles one
should worry about higher loop orders in the contributions to the
localized operators coming from the physics below $M_{s}$.  For these
strongly interacting particles, the multi-loop effects are of the same
order as the one-loop result.  Note, however, that for 
particles that do not interact directly with colored states, such as the 
leptons, the one-loop computation should be a good
approximation for the coefficient of the localized operator, 
modulo the bare contributions
which are not logarithmically enhanced.  At any rate, the above
results should be used carefully and we shall take them as an estimate
of the physics due to localized operators.  For the most part, we will
express our results in terms of generic localized parameters.
However, for numerical purposes we will use the above one-loop
expressions.

\subsection{Mass spectrum}
\label{masses}

The localized terms of Eqs.~(\ref{bterms}) and (\ref{gluon-terms})
shift the masses of the fermion, gauge field and spinless adjoint KK
modes, leading to mass splittings among the members of a given KK
level.
To lowest order in the localized terms $C_{pf}$, $C^{\prime}_{pf}$,
$C_{pA}$ and $C'_{pA}$, where $f$ stands for any of the fermions and
$A$ for any of the gauge fields, the mass shifts are
\bear
M_{f^{(j,k)}} &=& M_{j,k} \left( 1 - \frac{1}{2} \ZZ_{f^{(j,k)}} + 
\frac{1}{2} \ZZ^{\prime}_{f^{(j,k)}} \right) ~, \nonumber \\ [0.3em]
M_{A^{(j,k)}} &=& M_{j,k} \left( 1 - \frac{1}{2} \ZZ_{A^{(j,k)}} 
\right)~, 
\label{mass-shifts} \\ [0.3em]
M_{A^{(j,k)}_{H}} &=& M_{j,k} \left( 1 + \frac{1}{2} \ZZ_{A^{(j,k)}_{H}}
\right)~. 
\nonumber 
\eear
For KK-parity even fields, that is (j,k) modes with $(-1)^{j+k} = +1$, we
find 
\bear
\ZZ_{f^{(j,k)}} &=& \frac{1}{(\pi M_{s}R)^{2}}\, \left(2 C_{1f} + 
C_{2f}\right) ~, 
\nonumber \\[0.4em]
\ZZ^{\prime}_{f^{(j,k)}} &=& \frac{2}{(\pi M_{s} R)^{2}} \,{\rm Re} 
\left(2 C^{\prime}_{1f} + C^{\prime}_{2f}\right) ~, 
\nonumber \\[0.4em]
\ZZ_{A^{(j,k)}} &=& \frac{1}{(\pi M_{s}R)^{2}}\left(2 C_{1A} + 
C_{2A}\right) ~,  
\nonumber \\[0.4em]
\ZZ_{A^{(j,k)}_{H}} &=& \frac{1}{(\pi M_{s}R)^{2}}\left(2 C'_{1A} + 
C'_{2A}\right) ~,
\label{deltaZmass} 
\eear
while for KK-parity odd
fields, {\rm i.e.}, $(-1)^{j+k} = -1$, the 
$\ZZ$'s are given by (\ref{deltaZmass})
with $C_{2f} = C'_{2f} = C_{2A} = C'_{2A} = 0$. The mass shifts depend
on the quantum numbers $j$ and $k$ because the coefficients
$C_{pf}$, $C^{\prime}_{pf}$, $C_{pA}$ and $C'_{pA}$ are running
parameters and should be evaluated at the scale of the corresponding
mass $M_{j,k}$.

The mass of the gluon vector mode, $G_\mu^{(j,k)}$, can
be parametrized as
\be
M_{G^{(j,k)}} =  M_{j,k}\left( 1 + A_G \, C^{G}_{j,k} \right) ~,
\label{G-mass}
\ee
where
\be
C^{G}_{j,k} \equiv
\frac{g^2_s N_c}{16 \pi^2} 
\ln \left(\frac{M_s^2}{M^{2}_{j,k}}\right) ~,
\label{c-coefficient}
\ee
$M_{j,k}$ are the masses due to motion in the extra dimensions, given
in Eq.~(\ref{kk-mass}), and $A_G$ is a dimensionless parameter
expected to be of order unity.
The $SU(2)_W$-doublet quark modes have masses
\be
M_{Q^{(j,k)}_+} =  M_{j,k} \left(1 + A_{Q_+} C^{G}_{j,k}  
+ \frac{m_q^2}{2M_{j,k}^2} 
\right) ~,
\label{Q-mass}
\ee
where $m_q$ is the mass of the zero-mode quark, and we expanded in
$m_q^2/M_{j,k}^2 \ll 1$. We employ a similar
parametrization for the $SU(2)_W$-singlet quarks in terms of
dimensionless parameters, $A_{U_-}$ and $A_{D_-}$ (collectively denoted by
$A_{Q_-}$).
The coefficients $A_{Q_+}$, $A_{U_-}$ and $A_{D_-}$ are also expected to be
of order unity.

The masses of the hypercharge and electrically-neutral $SU(2)_W$
vector modes, $B_\mu^{(j,k)}$ and $W^{(j,k)3}_\mu$, are given by
\bear
M_{W^{(j,k)}} &=& M_{j,k}\left( 1 + \frac{2g^2}{N_c g_s^2} 
A_W \, C^G_{j,k} + \frac{M_W^2}{2 M_{j,k}^2 } \right) ~,
\nonumber \\ [0.6 em]
M_{B^{(j,k)}} &=& M_{j,k} \left( 1 + \frac{g^{\prime 2}}{N_c g_s^2} 
A_B \, C^{G}_{j,k} \right) ~,
\label{mwb}
\eear
where $g$ and $g^\prime$ are the 4D $SU(2)_{W}$ and $U(1)_Y$ gauge
couplings, respectively.  In Eqs.~(\ref{mwb}), we have neglected terms
of order $(M_W/M_{j,k})^4$, where $M_W$ is the zero-mode $W$ mass.  
Similar parametrizations can be used for
the masses of the spinless adjoints:
\bear
M_{G^{(j,k)}_{H}} &=& M_{j,k}\left( 1 + A_{G_{H}} \, C^G_{j,k} \right) ~,
\nonumber \\ [0.6 em]
M_{W^{(j,k)}_{H}} &=& M_{j,k}\left( 1 + \frac{2g^2}{N_c g_s^2} 
A_{W_{H}} \, C^G_{j,k} + \frac{M_W^2}{2 M_{j,k}^2 } \right) ~,
\nonumber  \\ [0.6 em]
M_{B^{(j,k)}_{H}} &=& M_{j,k} \left( 1 + \frac{g^{\prime 2}}{N_c g_s^2} 
A_{B_{H}} \, C^{G}_{j,k} \right) ~.
\label{mSpinless}
\eear
The $SU(2)_W$-doublet and -singlet lepton modes, 
$L^{(j,k)}$ and $E^{(j,k)}$, have masses
\bear
M_{L^{(j,k)}} &=& M_{j,k} \left(1 + \frac{2g^2}{N_c g_s^2} 
A_L \, C^G_{j,k}  \right) ~,
\nonumber \\ [0.5 em]
M_{E^{(j,k)}} &=& M_{j,k} \left( 1 + \frac{g^{\prime 2}}{N_c g_s^2} 
A_E \, C^G_{j,k} \right) ~.
\label{L-mass}
\eear
The above corrections to the KK masses of leptons and electroweak bosons
are due to the 6D $SU(2)_W\times U(1)_Y$ interactions.  Given that the
loop factor $C^G_{j,k}$ is computed for QCD, we have factored out the
electroweak gauge couplings such that the coefficients $A_W, A_B,
A_{W_{H}}, A_{B_{H}}, A_L$, and $A_E$ are all expected to be of
order unity, barring enhancement factors due to particle
multiplicities.

The KK modes of the Higgs boson, $h^{(j,k)}$,
are split in mass from the KK modes of the other three 
degrees of freedom of the Higgs doublet.
The latter ones mix with the $W^{(j,k)}_H$ KK modes,
giving rise to the Nambu-Goldstone bosons eaten
by $W^{(j,k)}_\mu$, and to the orthogonal states
$\tilde{\eta}^{(j,k)a}$ ($a=1,2,3$) which form a tower of 
physical spin-0 particles.
For a detailed discussion of this mechanism, we refer the reader
to Section 6 of Ref.~\cite{Burdman:2005sr}.
The masses of these Higgs KK modes may be parametrized as
\bear \hspace*{-2em}
M_{h^{(j,k)}} & = & 
M_{j,k}
\left( 1 + \frac{2g^2}{N_c g_s^2} A_H \, C^G_{j,k} 
+ \frac{M_h^2}{2 M_{j,k}^2}\right)~,
\nonumber \\ [0.6em] \hspace*{-2em}
 M_{\tilde{\eta}^{(j,k)}} & = & M_{j,k}\left( 1 + \frac{2g^2}{N_c g_s^2} 
A_{\tilde{\eta}} \, C^G_{j,k} + \frac{M_W^2}{2 M_{j,k}^2 } \right) ~,
\label{H-mass}
\eear
%
where we assumed that the Higgs boson mass $M_h$
is small enough compared to the compactification scale such that
the $(M_h/M_{j,k})^4$ corrections may be ignored.
In the limit $M_h \ll 1/R$, the KK modes of the Higgs boson,
$h^{(j,k)}$, and of the three eaten  Nambu-Goldstone bosons,
$\tilde{\eta}^{(j,k)\pm}$ and $\tilde{\eta}^{(j,k)3}$, 
form a degenerate 
$SU(2)_W$ doublet at each KK level, which we denote by $H^{(j,k)}$.

In this paper we are interested in the KK-parity even states, which
can be singly produced at colliders, as we will see in the next
section.  It will be useful to have the mass
shift formulae (\ref{mass-shifts}) and
(\ref{deltaZmass}) that follow from the one-loop results,
Eqs.~(\ref{rf})--(\ref{rA}), applied to the standard model gauge group
and field content. For the gluon vector modes, and 
for the $SU(2)_W$-doublet and -singlet quark modes,
$Q_+$ and $Q_-$ respectively, we find 
\bear
&& A_G=\frac{13}{3} ~,
\nonumber \\ [0.5em] 
&& A_{Q_+}= \frac{20}{9} + \frac{1}{4g_s^2} \left( \lambda_{q_L}^2
+ 5 g^2 + \frac{5}{27}g^{\prime\,2}  \right)~,
\nonumber \\ [0.5em] 
&& A_{Q_-}= \frac{20}{9} + \frac{1}{g_s^2} \left(\frac{1}{2} 
\lambda_{q_R}^2
+ \frac{5}{12} y_{q_R}^2 g^{\prime\,2}  \right)~,
\label{aq}
\eear
where $y_f$ are the hypercharges of the quarks and leptons, normalized
such that the quark doublets have $y=1/3$.  Here $\lambda_{q_L}$ and
$\lambda_{q_R}$ are the Yukawa couplings to the Higgs doublet (given
by $\lambda_{b_L} = \lambda_{t_L} = \lambda_{t_R} \equiv \lambda_t
\simeq 1$, and negligible for the other flavors).  Note that the top
Yukawa gives a positive contribution so that the third generation
$Q^{3}$ and $U^{3}$ KK modes are heavier than those of the first two
generations.  The positive contribution to the mass shifts due to
Yukawa couplings is special to six dimensions, and is related to the
existence of two 6D chiralities, both of which must be involved in the
Yukawa interaction (notice that in 5D the Yukawa couplings give a
negative contribution to the mass shifts \cite{Cheng:2002iz}).  Note
also that the KK gluons are heavier than the KK quarks.  However, as
we stressed before, for strongly interacting particles the one-loop
results should be taken only as indicative of the size of the mass
shifts.  Although a situation where the KK quarks are heavier than the
KK gluons is possible, we will assume that higher order
contributions do not change the hierarchy of masses found at one-loop.

For the electroweak gauge bosons, we get 
\be
A_W = \frac{85}{24}~,
\hspace{5mm} A_B=-\frac{83}{12} ~, 
\ee
while the leptons have
\bear
&& A_{L}= \frac{15}{8}\left(1 + \frac{ g^{\prime\,2}}{3 g^{2}}\right)~,
\nonumber \\ [0.5em] 
&& A_{E}= 5~,
\label{al-ae}
\eear
so that the $SU(2)_W$ gauge boson modes are heavier than the 
lepton modes.

\begin{figure}
\psfrag{mass}[B]{\hspace*{3.5em} Mass \ [GeV]}
\hspace*{-0.6em}
\psfig{file=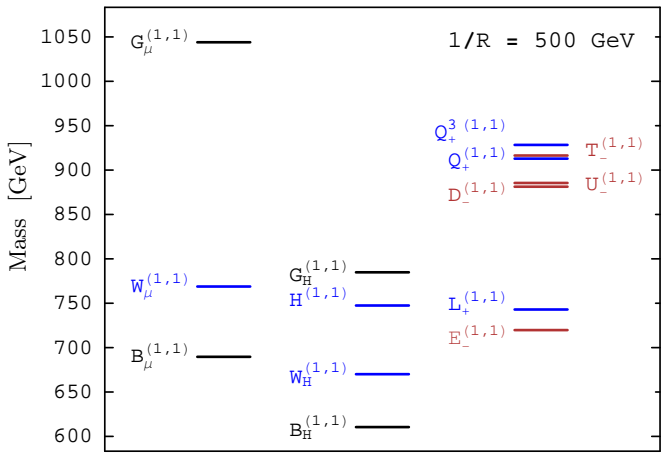,width=8.7cm,angle=0}
\vspace*{-8mm}
\caption{Mass spectrum of the (1,1) level for $1/R$ = 500 GeV.
Electroweak symmetry breaking effects are small, and have not 
been included.}
\label{fig:second}
\end{figure}

For the spinless adjoints the mass shifts arise from the
second term in Eq.~(\ref{gluon-terms}) and the analogous terms in the
electroweak sector.  As shown in \cite{Burdman:2005sr}, the
KK-expansion of the extra-dimensional field strength, $F_{45}$,
defines gauge invariant linear combinations of $A_{4}$ and $A_{5}$
that are orthogonal to the would-be Nambu-Goldstone bosons eaten by the
vector modes at each KK level.  Thus, only these gauge invariant
degrees of freedom, that we call spinless adjoints, get a mass shift
from the localized terms, given in the third equation
of~(\ref{mass-shifts}).  We obtain the following values 
for the parameters defined by Eqs.~(\ref{mSpinless}):
\be
A_{G_{H}} = 1~,
\hspace{5mm}
A_{W_{H}} = -\frac{17}{8}~,
\hspace{5mm}
A_{B_{H}} = -\frac{153}{4}~.
\label{spinless-mass}
\ee
Note that the $(1,1)$ $SU(3)_C$ spinless adjoints receive a positive
contribution to their masses, but are typically lighter than the
$(1,1)$ quarks.  Similarly, the electroweak spinless adjoints are
lighter than the $(1,1)$ leptons. Their masses are driven down by 
the contribution due to the fermions. 

Finally, the parameters that control the KK Higgs masses
in Eq.~(\ref{H-mass}) are given by 
\be
A_H \simeq A_{\tilde{\eta}} \simeq \frac{33}{32} 
+ \frac{\lambda_t^2}{2 g^2} ~, 
\ee
where we have not included the contributions from 
Higgs self-interactions and from $U(1)_Y$ interactions.

The mass spectrum of the 
$(1,1)$ modes is shown in Figure~\ref{fig:second} for $1/R = 500$ GeV.
Higher-loop contributions involving colored KK modes may be 
important (see the end of Section~\ref{sec:local}), and  
may shift the mass spectrum. This uncertainty is larger than 
corrections coming from the running of the coupling constants,
or electroweak symmetry breaking. 
We ignored these effects in  Figure~\ref{fig:second}, and 
we used some rough estimates for the couplings at the scale
$M_{1,1} = \sqrt{2}/R$: $(g/g_s)^2=0.34$,
$(g^\prime/g_s)^2=0.10$, $(\lambda_t/g_s)^2=0.8$,
$C^G =0.1$.
We also assumed that the Higgs boson is much lighter than the 
compactification scale.

\begin{figure}
\psfrag{mass}[B]{\hspace*{3.5em} Mass \ [GeV]}
\hspace*{-0.6em}
\psfig{file=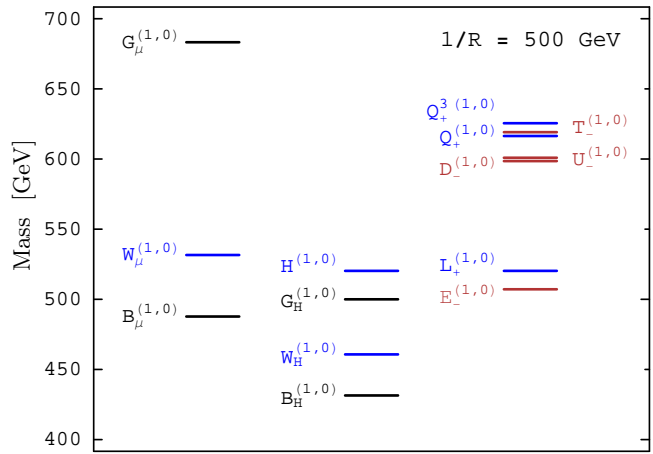,width=8.5cm,angle=0}
\vspace*{-4mm}
\caption{Mass spectrum of the (1,0) level. The lightest KK particle
is the $B_H^{(1,0)}$ spinless adjoint.}
\label{fig:first}
\end{figure}

We also point out here that at the $(1,0)$ level, the mass corrections to
the electroweak spinless adjoints are also negative.  The mass correction
to the $(1,0)$ $SU(3)_C$ spinless adjoints happens to vanish at one-loop
for the standard model field content, but one should keep in mind that
multi-loop contributions are expected to be important for the strongly
interacting particles.  The corresponding mass shifts for the spin-1 particles 
are positive for the $(1,0)$ gluons, and negative for the $(1,0)$ $W$ and 
$B$ vector modes. In fact, it is interesting that the lightest KK particle 
is predicted to be the spinless hypercharge mode, $B_{H}^{(1,0)}$. 
Thus, in contrast to the case of five dimensions, 
the natural dark matter candidate has spin-0.
The mass spectrum of the $(1,0)$ modes is shown in Figure~\ref{fig:first}
for $1/R = 500$ GeV.

\subsection{KK-number violating interactions}
\label{sec:KKcouplings}

The $Z_2^{\rm KK}$ symmetry implies that for any interaction among KK
modes the sum over all $j$ and $k$ numbers should be even.  In
particular, interactions involving two zero modes and a $(j,k)$ mode
with $j\geq 1$ and $j+k$ even is allowed.  Such an interaction is not
generated at tree level by bulk interactions, but arises due to the
localized operators.

To be concrete, the effective 4D, KK-number violating couplings
between zero-mode quarks and massive KK gluons are given by
\be
g_s C^{qG}_{j,k} \left( \overline{q} \gamma^\mu T^a q\right)  
G_\mu^{(j,k)a} ~,
\label{coupling}
\ee
where $C^{qG}_{j,k}$ are real dimensionless parameters, $T^a$ are the
$SU(3)_C$ generators in the fundamental representation, $g_s$ is the QCD
gauge coupling, and $q$ stands for any of the standard model quarks.
The strength of the couplings to zero-mode fermions is controlled by
the kinetic terms localized at the fixed points, and contained in
${\cal L}_1$ and ${\cal L}_2$ in Eq.~(\ref{l4d}).  Their dimensionless
coefficients ultimately determine the strength of the KK-number
violating couplings of fermions to gauge bosons.  In terms of the
coefficients defined in Eqs.~(\ref{bterms}) and (\ref{gluon-terms}),
$C^{qG}_{j,k} $ is given to lowest order in the localized terms by
\bear
C^{qG}_{j,k} = - \frac{1}{2} \, \overline{\ZZ}_{G^{(j,k)}} + 
\frac{1}{2} \, \overline{\ZZ}_{q^{(j,k)}} - 
\frac{1}{2} \, \overline{\ZZ}^{\prime}_{q^{(j,k)}} ~,
\label{interactions}
\eear
where it is understood that $G^{(j,k)}$ is KK-parity even [i.e. $j+k$
is even], but not the zero-mode, and
\bear
\overline{\ZZ}_{q^{(j,k)}} &=& \frac{1}{(\pi M_{s}R)^{2}}
\left(2 C_{1q} + (-1)^{j} C_{2q}\right) ~, \nonumber \\ [0.4em]
\overline{\ZZ}^{\prime}_{q^{(j,k)}} &=& \frac{2}{(\pi M_{s}R)^{2}}
\,{\rm Re}\left(2 C^{\prime}_{1q} + (-1)^{j} C^{\prime}_{2q} \right)~, 
\nonumber \\[0.4em]
\overline{\ZZ}_{G^{(j,k)}} &=& \frac{1}{(\pi M_{s}R)^{2}}
\left(2  C_{1G} + (-1)^{j} C_{2G}\right) ~. 
\label{deltaZcoupling} 
\eear
Notice that when both $j$ and $k$ are even we can write
Eq.~(\ref{interactions}) in terms of the mass shifts given in
Eq.~(\ref{mass-shifts}) as
\be
C^{qG}_{j,k} = \frac{\delta M_{G^{(j,k)}}}{M_{j,k}} - 
\frac{\delta M_{q^{(j,k)}}}{M_{j,k}} \;\; , \;\; j,k \; {\rm even}  ~.
\label{even-couplings}
\ee
However, when both $j$ and $k$ are odd the above relation does not
hold (unless there are no terms induced at $(x^{4},x^{5}) = (0,\pi
R)$, i.e. $C_{1q} = C^{\prime}_{1q} = C_{1G} = 0$).  In this case, the
above interactions depend on a different combination of the
coefficients in ${\cal{L}}_{1}$ and ${\cal{L}}_{2}$ than the KK mass
shifts and are, effectively, independent parameters.

\medskip

As mentioned above, the coefficients of the localized operators
receive contributions from physics at the cut-off $M_s$, and run
logarithmically below $M_s$ due to bulk loop
effects.  The contribution to the
localized couplings from physics between the scales $M_s$ and $\mu <
M_s$ is of the order of $(g^2 N_c/16\pi^{2}) \ln (M^2_s/\mu^2)$.  This
contribution is enhanced compared to the bare one by a logarithmic
factor.  Based on these NDA estimates, Eqs.~(\ref{interactions}) and
(\ref{deltaZcoupling}) give the values of the parameters
$C^{qG}_{j,k}$ at the scale of the KK mode mass $M_{j,k}$:
\be
\label{xi-def}
C^{qG}_{j,k} = \xi^G_q \, C^{G}_{j,k} ~,
\ee
where $\xi^G_q$ is a dimensionless parameter of order unity, and
$C^{G}_{j,k}$ was defined in Eq.~(\ref{c-coefficient}).

The localized operators at the cut-off scale may be flavor dependent.
Their contributions to the $C^{qG}$ coefficients are not shown in
Eq.~(\ref{c-coefficient}) because they are not enhanced by the
logarithmic factor.  Nevertheless, Eq.~(\ref{separation}) implies that
the logarithmic factor is at most as large as $\ln (M_sR)^2 \approx
4.6$, and therefore the flavor dependent operators may lead to large
flavor-changing neutral processes.  In order to suppress these, the
physics above the cut-off scale must possess some approximate flavor
symmetry.

The KK spinless adjoints interact with the zero-mode quarks only via
dimension-5 or higher operators:
\be
\frac{g_s \tilde{C}^{qG}_{j,k}}{M_{j,k}} \, 
\left(\overline{q} \gamma^\mu T^a q \right)\, D_\mu G_H^{(j,k)a}  
~,
\label{GH-coupling}
\ee
where $D_\mu$ is the gauge covariant derivative, and 
$\tilde{C}^{qG}_{j,k}$ are real dimensionless parameters
that depend on the flavor and {\it chirality} of the quark $q$.
Note that $G_H$ has axial couplings to quarks, proportional to
the difference between the coefficients of the 
left- and right-handed quark operators.
The largest contributions to these coefficients arise from the quark and
spinless adjoint kinetic and mass mixing effects associated with the
localized kinetic terms of Eqs.~(\ref{bterms}) and
(\ref{gluon-terms}).  There is also a subdominant finite direct vertex
contribution, suppressed by a logarithm compared to the mixing
effects.  According to NDA, the direct bare contributions from
localized operators to the interaction (\ref{GH-coupling}) are of
order $(g^{2}_{s} N_{c}/16\pi^{2})^{2}$, and are therefore negligible
compared to the contributions due to mixing effects.  It is important
to notice that the vertex (\ref{GH-coupling}) is proportional to the
quark masses, as can be seen by integrating by parts and using the
fermion equations of motion.  As a result, the spinless adjoints decay
almost exclusively into top quarks.  This observation also implies
that the coupling for direct production of the spinless states is
negligible, being suppressed by the $u$ or $d$-quark masses.  However,
the spinless adjoints can be easily produced in the decays of KK
quarks or leptons, as we discuss in the next section.

Dimension-4 operators which couple a single spinless adjoint to one or two 
zero-mode gluons are forbidden by the unbroken gauge invariance
associated with the zero-mode gauge boson.
Naively, a spinless adjoint may couple to zero-mode gluons via
the dimension-5 operator
\begin{equation}
\frac{1}{M_{j,k}} {\rm Tr}\left(G^{\mu\nu} G_{\mu\nu} G_H^{(j,k)}\right)
\label{thisiszero}
\end{equation}
or the analogous operator involving a dual field strength.
However, these operators vanish 
because the trace over three generators $T^a, T^b, T^c$ in 
the adjoint representation is proportional to the antisymmetric 
structure constant $f^{a b c}$, and two indices are contracted with identical 
gluon field strengths. On the other hand, the operator
obtained by replacing one of the gluon field strengths 
in (\ref{thisiszero}) by the field strength of a KK gluon need not vanish, 
and generates a coupling of a spinless adjoint to a KK gluon and a gluon
zero mode.

Higher-dimension
operators coupling the hypercharge spinless adjoint
to two spin-1 fields, {\it e.g.},
${\rm Tr}(G^{\mu\nu} G_{\mu\nu} B_H^{(j,k)})$  or
$B^{\mu\nu} B_{\mu\nu} B_H^{(j,k)}$,
could be present.
Interestingly, the one-loop contributions to their coefficients vanish. This is because
$B_H^{(j,k)}$ couples to fermions via an axial-scalar coupling so that the relevant triangle diagrams are proportional to the corresponding gauge anomaly coefficients, which are canceled.
%
%

\medskip

The $W^{(j,k)a}_\mu$ KK modes couple to zero-mode fermions through
\be
g \, C^{fW}_{j,k} \, 
\left( \overline{f}_L \gamma^\mu T^a_2 f_L \right) W_\mu^{(j,k)a} ~,
\label{coupling-W}
\ee
where $T^{a}_2$ are the $SU(2)_W$ generators, $g$ is the 4D $SU(2)_W$
gauge coupling, and $f$ are the quark and lepton fields of any
generation.  It is convenient to write the dimensionless parameters
$C^{fW}_{j,k}$ as follows:
\bear
\label{xiw-def}
C^{qW}_{j,k} &\equiv & \xi^W_q \, C^{G}_{j,k} ~, 
\nonumber \\ [0.5em] 
C^{lW}_{j,k} &\equiv & 
\frac{2 g^2}{g_s^2 N_c} \, \xi^W_l \, C^{G}_{j,k} ~.
\eear
Here, $l$ stands for the $SU(2)_W$-doublet leptons.  The $\xi^W_q$ and
$\xi^W_l$ parameters are estimated via NDA to be of order unity.

Similarly, the $B^{(j,k)}_\mu$ KK modes couple to zero-mode quarks and
leptons:
\be
g^\prime \frac{y_f}{2} \, C^{fB}_{j,k} \, 
\left( \overline{f} \gamma^\mu f \right) B_\mu^{(j,k)} ~,
\label{coupling-B}
\ee
where $g^\prime$ is the 4D $U(1)_Y$ gauge coupling, and $y_f$ are the
hypercharges of the quarks and leptons, normalized such that the quark
doublets have $y=1/3$.  The $C^{fB}_{j,k}$ parameters may be written
in terms of other parameters ($\xi^B_q$, $\xi^B_l$, $\xi^B_e$)
expected to be of order unity:
\bear
\label{xib-def}
C^{qB}_{j,k} &\equiv & \xi^B_q \, C^{G}_{j,k} ~, 
\nonumber \\ [0.5em] 
C^{lB}_{j,k} &\equiv & 
\frac{2 g^2}{g_s^2 N_c} \,\xi^B_l \, C^{G}_{j,k} ~,
\nonumber \\ [0.5em] 
C^{eB}_{j,k} &\equiv &
\frac{g^{\prime 2}}{g_s^2 N_c} \,\xi^B_e\, C^{G}_{j,k} ~.
\eear
%


So far we have parametrized the (0,0)(0,0)(j,k) couplings, using NDA
as a guide to argue that certain parameters are expected to be of
order unity.
The explicit one-loop results, Eqs.~(\ref{rf})--(\ref{rA}), involve a
$(\,(j + k)/2,(k - j)/2\,)$ mode in the loop when $j \geq k$, and a
$(\,(j - k)/2,(j + k)/2\,)$ mode when $j < k$ \cite{Dobrescu:2004zi}.
For $j$ and $k$ even, there is also a one-loop contribution with a
$(j/2,k/2)$ mode running in the loop.  As a result, the
$(0,0)(0,0)(1,1)$ interaction is generated by a $(1,0)$ loop, while
the $(0,0)(0,0)(2,0)$ interaction is generated by the sum of a $(1,0)$
loop and a $(1,1)$ loop.
As we will see below, these interactions of KK modes beyond the
$(1,0)$ level with two zero-mode fields, have important
phenomenological consequences.  Also note that the leading
contributions to these one-loop diagrams involve gluon and quark KK
modes, and therefore are flavor independent.  Electroweak KK modes
induce some splitting between the couplings of the $SU(2)_W$-doublet
and singlet quarks as well as between the up- and down-type quarks.
Another effect is due to the Yukawa couplings to the Higgs doublet, 
and is notable only for the $(t_L,b_L)$ and $t_R$ quark fields.

If we use Eqs.~(\ref{rf})--(\ref{rA}) we find, in the case of 6D gauge
fields $A_\alpha$ interacting with some 6D Weyl fermions $f$ and 6D
complex scalars $s$ which have zero modes, that the
``renormalization'' constants (\ref{deltaZcoupling}) that determine
the couplings of the $(1,1)$ vector modes to zero-mode fermions via
Eq.~(\ref{interactions}), are given by
\bear
\overline{\ZZ}_{f^{(1,1)}} &=& \left[ -6 \sum_A C_2(f) \, g_A^2 + 
\sum_i \lambda_i^2 \right]
\nonumber \\ [0.5em] 
&& \times
\frac{1}{16\pi^2} \ln \left( \frac{M_s^2 R^2}{2} \right)  ~,
\nonumber \\ [0.7em] 
\overline{\ZZ}_{f^{(1,1)}}^\prime  &=&  \sum_i \lambda_i^2
\frac{1}{8\pi^2} \ln \left( \frac{M_s^2 R^2}{2} \right) ~,
\nonumber \\ [0.6em] 
\overline{\ZZ}_{A^{(1,1)}} &=& \left[-\frac{22}{3}C_2(A) + \frac{4}{3} 
\sum_f T(f) 
+ \frac{2}{3} \sum_s T(s) \right] g_A^2 
\nonumber \\ [0.5em] 
&& \times
\frac{1}{16\pi^2} \ln \left( \frac{M_s^2 R^2}{2} \right)  ~.
\label{}
\eear

Applying this result to the standard model gauge group, we find the parameters
introduced in Eq.~(\ref{xi-def}),
\bear
\label{xig-predict}
&& \xi_{q_L}^G = 1  - \frac{1}{2g_s^2} \left( \frac{1}{3} 
\lambda_{q_L}^2
+ \frac{3}{2} g^2 + \frac{1}{18}g^{\prime\,2}  \right)~,
\nonumber \\ [0.5em] 
&& \xi_{q_R}^G = 1 - \frac{1}{g_s^2} \left(\frac{1}{3} \lambda_{q_R}^2
+ \frac{y_{q_R}^2}{4} g^{\prime\,2} \right) ~.
\eear
Similarly, the parameters that control the $W^{(1,1)}_\mu$ couplings
to zero-mode fermions are given by
\bear
\label{xiw-predict}
&& \xi_{q}^W = - \frac{4}{3} - \frac{\lambda_q^2}{6g_s^2}
+ \frac{1}{36g_s^2} \left(11g^2 - g^{\prime\,2} \right) ~, 
\nonumber \\ [0.5em] 
&& \xi_{l}^W = \frac{11}{24} - \frac{3g^{\prime\,2}}{8g^2} ~.
\eear
Note that in the above equations the hypercharge interaction give only
small corrections, which for practical purposes may be neglected in
what follows.  However, in the case of the $B^{(1,1)}_\mu$ couplings
to zero-mode fermions, the wavefunction renormalization of
$B^{(1,1)}_\mu$ (more precisely, its mixing with $B^{(0,0)}_\mu$) is
enhanced by the large number of fields, giving a relatively large
contribution from the hypercharge interaction:
\bear
\label{xib-predict}
&& \xi_{q_L}^B = - \frac{4}{3}  - \frac{1}{2g_s^2} 
\left( \frac{1}{3} \lambda_{q_L}^2 + \frac{3}{2}g^2 + 
\frac{83}{18}g^{\prime\,2} \right)  ~, 
\nonumber \\ [0.5em] 
&& \xi_{q_R}^B =  - \frac{4}{3} - \frac{1}{g_s^2} \left[ \frac{1}{3} 
\lambda_{q_R}^2 
+ \left(\frac{41}{18}+\frac{y^2_{q_R}}{4}\right)g^{\prime\,2}  \right]  
~,   
\nonumber \\ [0.5em] 
&& \xi_{l}^B =  - \frac{9}{8}  - \frac{91g^{\prime\,2}}{24g^2}  ~,   
\nonumber \\ [0.5em] 
&& \xi_{e}^B =  - \frac{59}{6} ~.
\eear
Note that the couplings of the spin-1 KK modes to the third generation
quarks are somewhat enhanced due to the loops involving Higgs KK
modes.

\section{Decays of the (1,1) modes}\label{sec:decays}
\setcounter{equation}{0}

The 6D Standard Model outlined in the previous section leads to
specific predictions for the properties of the KK modes.  In this
section we compute the branching fractions of the (1,1) modes, which
are of special interest for collider phenomenology.  We start by
summarizing the information regarding the parameters that control the
branching fractions, and then we discuss in turn each of the (1,1)
modes.

\subsection{Parameters}

The most important parameter is $1/R$, which sets the overall scale of
the KK mass spectrum.  Lower limits on $1/R$ are set mainly by 
electroweak precision constraints and are likely to be in the 
$300-500$~GeV range, based on the study from 
Refs~\cite{Appelquist:2000nn,Appelquist:2002wb},
where a different compactification of two extra dimensions has been 
considered. We emphasize that the limits from electroweak constraints
are sensitive to higher-dimensional operators at the cut-off scale,
and therefore are not nearly as robust as the limits that will be set
by collider searches (see Section \ref{sec:limit}).

The cut-off scale, $M_s$, is in principle another free parameter.
However, it cannot be too close to $1/R$ because the effective field
theory loses its validity, and it cannot be too far above $1/R$
because the effective field theory becomes non-perturbative.  We use
the NDA estimate given in Eq.~(\ref{separation}):
$M_s = 10/R$.  Given that the physical observables
depend on the cut-off scale only logarithmically, this choice is not a
source of large uncertainties.  

The coefficients of the localized 
kinetic terms are all parameters beyond those in the standard model.
They control the leading contributions to the mass splittings among
the modes within a given KK level, as well as the KK-number violating
couplings, as discussed in the previous section.  We note that there
are two parameters per localized operator: one controls the strength
of the operator at $(0,0)$ and $(\pi R,\pi R)$, while the second one
controls the strength at $(0,\pi R)$.  It is important to notice that
the mass shifts and KK-number violating couplings depend on different
linear combinations of these two parameters.  Since in this paper we
are mainly interested in the phenomenology of the $(1,1)$ level, we
may consider their masses and couplings as independent parameters.
However, one should keep in mind that knowledge of these parameters
imply definite relations for the masses and couplings of other states,
such as the $(1,0)$ modes.

The loop factor that controls the
couplings of the $(1,1)$ gluons to zero-mode quarks $C^{G}$ is (we drop the
$j=k=1$ indices)
\be
C^{G} =
\frac{\alpha_s N_c}{4 \pi} 
\ln \left(\frac{M_s^2R^2}{2}\right) \simeq 0.1 ~.
\label{c11-coefficient}
\ee
where we used the value of the strong coupling constant at a scale of
about 1 TeV, $\alpha_s = 0.1$.

The $G^{(1,1)}_\mu$, $W^{(1,1)3}_\mu$ and $B^{(1,1)}_\mu$ particles
have couplings to zero-mode fermions proportional to the parameters of
order one $\xi^G_q$, $\xi^W_q$, $\xi^W_l$, $\xi^B_q$, $\xi^B_l$, and
$\xi^B_e$, introduced in Eqs.~(\ref{xi-def}), (\ref{xiw-def}) and
(\ref{xib-def}).  We will keep the dependence on the $\xi$ parameters
explicit whenever possible, but for numerical results we use the
one-loop values given in Eqs.~(\ref{xig-predict}), (\ref{xiw-predict})
and (\ref{xib-predict}).  One should emphasize that the decay widths
and production cross sections depend quadratically on the $\xi$
parameters.  Therefore, the estimates for the $\xi$ parameters, which
for strongly interacting particles could be off by a factor as large
as 2 or so, is a major source of uncertainty in the predictions of
this model.  Barring unexpectedly large bare contributions, the $\xi$
parameters associated with the weakly interacting particles are
expected to be more reliable.  In addition, one should keep in mind
that the flavor independence of the (1,1) couplings may be only
approximate: the localized operators induced by physics at the cut-off
scale may be flavor dependent (as discussed in Section II, this is a
subdominant effect because the coefficients of these operators are not
enhanced by a logarithmic factor).

Other parameters control the mass splittings among various (1,1)
states.  These are also determined by the coefficients of the
localized operators, so that they are related to $C^{G}$ as shown in
Eqs.~(\ref{G-mass})-(\ref{H-mass}).  We will keep the dependence on
the coefficients $A_G$, $A_{G_{H}}$, $A_{Q_+}$, $A_{Q_-}$, 
$A_W$, $A_{W_{H}}$, $A_L$,
$A_B$, $A_{B_{H}}$, $A_E$ and $A_{H}$ explicit in our analytic
results, while in the numerical analysis we will use the values
given in Eqs.~(\ref{aq})-(\ref{spinless-mass}).

\subsection{Branching fractions of the $B_\mu$ (1,1)-mode} 
\label{sec:BDecays}

As we discussed in subsection~\ref{masses}, $B^{(1,1)}_\mu$ is the lightest
of the standard model KK excitations at the $(1,1)$ level.  
Only its spinless partner,
$B^{(1,1)}_H$, is expected to be lighter.  Thus, $B^{(1,1)}_\mu$ can
only decay into zero modes or into $B^{(1,1)}_H$ plus zero modes.  We
consider first the decays into zero modes only.  Their widths
may be computed in terms of the couplings given in
Eqs.~(\ref{coupling-B}).  The decay width into $q\bar{q}$, with
$q=u,d,s,c$, is given by
\be
\Gamma\left(B^{(1,1)}_\mu \to q\overline{q}\right)= \Gamma_0^B
\left[ \rule{0mm}{3.5mm}
(\xi^B_{q_L})^2 + 8 (\xi^B_{u_R})^2 + 2 (\xi^B_{d_R})^2 \right] ~,
\ee
where we summed over the four $q\bar{q}$ flavors, and we defined
\be
\Gamma_0^B \equiv
\frac{\alpha}{18 \cos^2\theta_{w}} \left(C^G\right)^2 M_{B^{(1,1)}} ~.
\ee
Here $\theta_w$ is the weak mixing angle, and $\alpha$ is the
electromagnetic coupling constant at a scale of order $1/R$.  The
decay widths into $t\bar{t}$ and $b\bar{b}$ are as follows:
\bear
\Gamma\left(B^{(1,1)}_\mu \to t \bar{t}\right)& = & \Gamma_0^B
\left[ \frac{1}{4}(\xi^B_{t_L})^2 + 4 (\xi^B_{t_R})^2 \right] 
\nonumber \\ [0.5em]
 &\times&\!\! \left(1 - \frac{m_t^2}{M_{B^{(1,1)}}^2} \right) 
\left(1 - \frac{4m_t^2}{M_{B^{(1,1)}}^2} \right)^{\!1/2} 
~,
\nonumber \\ [1em]
\Gamma\left(B^{(1,1)}_\mu \to b \bar{b}\right) & = & \Gamma_0^B
\left[ \frac{1}{4}(\xi^B_{t_L})^2 + (\xi^B_{d_R})^2 \right] ~.
\label{btt}
\eear
Note that we have neglected 
the QCD corrections and the $b$-quark mass.

The leptonic decays of $B_\mu^{(1,1)}$ are induced by the 6D
electroweak interactions:
\bear
\Gamma\left(B^{(1,1)}_\mu \to \ell^+\ell^-\right) & = &
\Gamma_0^B\frac{\alpha^2}{3\alpha_s^2 \sin^4\!\theta_{w}}
 \nonumber\\ [0.5em]
&& \times \left[ \rule{0mm}{3.5mm}(\xi^B_l)^2 +  (\xi^B_e)^2 
\tan^4\!\theta_{w} \right] ~,
 \nonumber\\ [1em]
\Gamma\left(B^{(1,1)}_\mu \to \nu\overline{\nu}\right) & = & 
\Gamma_0^B\frac{\alpha^2}{\alpha_s^2 \sin^4\!\theta_{w}} 
(\xi^B_l)^2 ~,
\label{GammaBll}
\eear
where $\ell =e,\mu,\tau$, and there is no sum over the three charged
lepton pairs, while the decay width into $\nu\overline{\nu}$ is summed
over the three neutrino flavors.  

There are also 3-body decays,
$B^{(1,1)}_\mu \to B^{(1,1)}_{H} \ell^+\ell^-$ through an off-shell
(1,1) lepton. The decay that proceeds
through off-shell $SU(2)_W$-singlet leptons dominates 
because of their larger hypercharge.
In the limit that $B^{(1,1)}_\mu$ and $B^{(1,1)}_H$ are
almost degenerate, $M_{B} - M_{B_{H}} \ll M_{B}$, we estimate that 
each tree-level diagram (ignoring interference) contributes at most
\bear
& & \hspace{-4em} \Gamma\left(B^{(1,1)}_\mu \to B^{(1,1)}_{H} 
\ell^+\ell^-\right) 
\nonumber \\ [0.4em] 
&& \lae \frac{\alpha^{2}
M_{E^{(1,1)}}^2\left(M_{B^{(1,1)}} - M_{B_{H}^{(1,1)}}\right)^{4}}
{\pi \cos^{4}\!\theta_{w}\, M_{B^{(1,1)}}( M^{2}_{E^{(1,1)}} 
- M^{2}_{B_{H}^{(1,1)}})^{2}} 
\nonumber \\ [0.4em]
&& \approx \Gamma_0^B \, 
\frac{\alpha^{3}\left(A_{B} - A_{B_{H}}\right)^{4}}
{2\pi \alpha^{2}_{s} \cos^{6}\!\theta_{w}
\left(A_E - A_{B_{H}}\right)^{2}}~.
\hspace{2mm}
\label{GammaBBHll}
\eear
%
In the second
equality we used the parametrizations of the masses given in
section~\ref{masses}, and expanded to lowest order in the mass
shifts.  Although for the 1-loop values of $A_{B}$, $A_{B_{H}}$ and
$A_{L}$ given in section~\ref{masses} there is a considerable
``resonant'' enhancement, it is not enough to overcome the phase space
suppression and we find that this partial decay width is at least one
order of magnitude smaller than the two-body decay into leptons of
Eq.~(\ref{GammaBll}).  
Given that the two-body leptonic decay is much smaller than the 
two-body decay into $q\overline{q}$, we conclude that 
$B_\mu^{(1,1)}$ is almost ``leptophobic'' (this term was coined in
Ref.~\cite{Babu:1996vt}).

The decays of $B_\mu^{(1,1)}$ into $B_H^{(1,1)}$ and a $Z$ boson 
or a photon could proceed through higher-dimension operators similar to those discussed after Eq.~(\ref{thisiszero}).  Such effects are suppressed compared to the two-body decays
discussed above, and therefore we neglect them. Note, however, that the decays of the first-level vector mode $B_\mu^{(1,0)}$
through such higher-dimension operators could be
phenomenologically relevant.

The $B^{(1,1)}_\mu$ total width, in
the limit $(m_tR)^2\ll 1$ and neglecting~(\ref{GammaBBHll}),
is given approximately by
%
\bear
\Gamma_B & = & \Gamma_0^B \left\{ \rule{0mm}{4.5mm}
(\xi^B_{q_L})^2 + 8 (\xi^B_{u_R})^2 + 3 (\xi^B_{d_R})^2 
+ \, \frac{1}{2}(\xi^B_{t_L})^2 \right.
 \nonumber\\ [0.5em]
&+& \left.\!\!   4 (\xi^B_{t_R})^2  
+ \frac{\alpha^2}{\alpha_s^2}
 \left[ \rule{0mm}{3.5mm} \frac{2(\xi^B_l)^2}{\sin^4\!\theta_{w}} 
+  \frac{(\xi^B_e)^2}{\cos^4\!\theta_{w}} \right]
\rule{0mm}{4.5mm}\right\} ~.
\eear

We compute the $\xi$ parameters from Eqs.~(\ref{xib-predict}), using
$\alpha = 1/127$, $\sin^2\theta_{w} = 0.23$, $\alpha_s = 0.1$, and
$\lambda_t = 1$, which gives $(g/g_s)^2=0.34$,
$(g^\prime/g_s)^2=0.10$, and $(\lambda_t/g_s)^2=0.8$.  
Note that these values for the coupling constants
are our rough estimates of their average values at a scale $M_{1,1}$
in the range $\sim 0.4 - 1$ TeV.
In order to use more precise values for the coupling constants
one would need to compute the changes in their running due to all 
lighter particles, including the (1,0) modes. However, the
masses of the colored (1,0) modes may get relatively 
large corrections from localized operators at the cut-off scale, so that 
we cannot include a precise scale dependence of the coupling constants.

For quarks, the
resulting values of the $\xi$ parameters are of order unity, as expected:
$\xi^B_{q_L} \simeq -1.8 $, 
$\xi^B_{u_R} \simeq \xi^B_{d_R} \simeq -1.6 $,
$\xi^B_{t_L} \simeq - 2.0 $,
$\xi^B_{t_R} \simeq -1.9 $.
For leptons, the couplings are somewhat enhanced as discussed in
subsection~\ref{sec:KKcouplings}: 
$\xi^B_{l} \simeq - 2.3$, $\xi^B_e \simeq - 9.8 $.  
The branching fractions computed with these
parameters are 59\% into dijets (not including $b$ jets), 30\%
into $t\bar{t}$, 7.1\% into $b\bar{b}$, 0.9\% for each of the
$e^+e^-$, $\mu^+\mu^-$ and $\tau^+\tau^-$ pairs, and 1.1\% for
invisible decays.  The total width is $\Gamma_B \simeq 2.4 \times
10^{-4} M_{B^{(1,1)}}$, so $B^{(1,1)}_\mu$ is an extremely narrow
resonance.

For larger values of $m_t R$, the branching fraction
into  $t\bar{t}$ is reduced as in Eq.~(\ref{btt}), 
while the other branching fractions are somewhat increased 
(see the values in Table~\ref{tab:BRw} of section~\ref{sec:QDecays} 
for $1/R =500$ GeV).

\subsection{Branching fractions of the $W^3_\mu$ (1,1)-mode} 
\label{sec:WDecays}

The width of the $W^{(1,1)3}_\mu$ decay to quarks depends on the
couplings $\xi^W_q$ defined in Eq.~(\ref{xiw-def}):
\be
\Gamma(W^{(1,1)3}_\mu\to q\bar{q} ) = 4 \, \Gamma_0^W (\xi^W_q)^2 ~,
\ee
where $q=u,d,s,c$, we summed over these four $q\bar{q}$ flavors, and
we ignored QCD corrections.  Here we defined
\be
\Gamma_0^W \equiv
\frac{\alpha}{8 \sin^2\!\theta_{w}} \left(C^G\right)^2 M_{W^{(1,1)}} ~.
\ee
For $t\bar{t}$ and $b\bar{b}$ final states we find
\bear
\Gamma\left(W^{(1,1)3}_\mu \to t \bar{t}\right)& = & 
\Gamma_0^W (\xi^W_t)^2 \left(1 - \frac{m_t^2}{M_{W^{(1,1)}}^2} \right)
\nonumber \\ [0.6em]
&\times & \!\!
\left(1 - \frac{4 m_t^2}{M_{W^{(1,1)}}^2} \right)^{\!1/2} ~,
\nonumber \\ [1em]
\Gamma\left(W^{(1,1)3}_\mu \to b \bar{b}\right) & = & \Gamma_0^W
(\xi^W_t)^2 ~.
\eear
The leptonic widths of $W^{(1,1)3}_\mu$ are the same for each neutrino
flavor or left-handed charged lepton:
\be
\Gamma(W^{(1,1)3}_\mu\to e_L^+ e_L^- ) = \Gamma_0^W (\xi^W_l)^2
\frac{4\alpha^2}{27\alpha_s^2 \sin^4\!\theta_{w}} ~.
\ee

If the $W^{(1,1)3}_\mu$ boson is heavier than the (1,1) leptons, then
it may also decay into a (1,1) lepton and a zero-mode lepton.  In fact
this is the case for the values of the masses given in Section~II.
Summing over the decay widths into leptons and anti-lepton doublets of
the three generations, we obtain
\bear
\Gamma\left(W^{(1,1)3}_\mu\!\to \sum L^{(1,1)} l \right) & \!=\! &
\frac{\alpha M_{W^{(1,1)}} }{2 \sin^2\!\theta_{w}}
\left(1 - \frac{M_{L^{(1,1)}}^2}{M_{W^{(1,1)}}^2} \right)^{\! 2} 
\nonumber \\ [0.5em] 
& \times &\!\!\left(1 + \frac{M_{L^{(1,1)}}^2}{2 M_{W^{(1,1)}}^2}\right) 
~.
\eear
To first order in the mass shifts shown in Eqs.~(\ref{mwb}) and
(\ref{L-mass}), we find
\be
\Gamma\left(W^{(1,1)3}_\mu\to \sum L^{(1,1)} l \right) \simeq
\Gamma_0^W \,
\frac{32\alpha^2 (A_W - A_L)^2}{3\alpha_s^2 \sin^4\!\theta_{w}} ~.
\ee
Adding these decay modes to the ones into zero modes, we find the
total width of $W^{(1,1)3}_\mu$ in the $(m_tR)^2\ll 1$ limit:
\bear\hspace*{-6em}
\Gamma_W \!& \!= & \Gamma_0^W \left\{
4(\xi_q^W)^2 +  (\xi_t^W)^2 
\left(2 - \frac{3 m_t^2}{M_{W^{(1,1)}}^2} \right)\right.
\nonumber \\ [0.5em]
&+& \!\!
\left.\frac{8\alpha^2 }{9\alpha_s^2 \sin^4\!\theta_{w}} 
\left[ \rule{0mm}{3.5mm} (\xi_l^W)^2 + \; 12(A_W - A_L)^2 \right]
\rule{0mm}{6mm}\right\} ~.
\eear

Using the $\xi$ parameters from Eq.~(\ref{xiw-predict}), $\xi^W_q
\simeq -1.2$, $\xi^W_l \simeq 0.35 $ and $\xi^W_t \simeq -1.4 $, 
and the $A$ parameters from Eq.~(\ref{al-ae}),
$A_W-A_L\simeq 1.5$, 
we find the branching fractions into $t\overline{t}$,
\be
{\rm Br}\left(W^{(1,1)3}_\mu\!\to t\overline{t}\right) \simeq
15\%\left(1 - \frac{2.6 \,m_t^2}{M_{W^{(1,1)}}^2} \right) ~,
\label{wttbar}
\ee
and the decays that preserve KK number, 
\be
{\rm Br}\left(W^{(1,1)3}_\mu\!\to \sum L^{(1,1)} l \right) \simeq
22\%\left(1 + \frac{0.44 \,m_t^2}{M_{W^{(1,1)}}^2} \right) ~,
\ee
For $m_t^2 \ll M_{W^{(1,1)}}^2$, the $W^{(1,1)3}_\mu$
has branching fractions of $48\%$ into dijets (not including the $b$ quark), 
$15\%$ into $b\bar{b}$,
$0.02\%$ into each of the $e^+e^-$, $\mu^+\mu^-$ and $\tau^+\tau^-$
pairs, and $0.06\%$ for invisible decays. Including the 
next order in the $m_t^2/M_{W^{(1,1)}}^2$ expansion,
these branching fractions have the 
same $m_t$ dependence as in Eq.~(\ref{wttbar}).
The $W^{(1,1)3}_\mu$ is
almost as narrow as $B^{(1,1)}_\mu$, with a total width $\Gamma_W
\simeq 10^{-3} M_{W^{(1,1)}}$.

\subsection{Quark and lepton (1,1)-mode branching fractions}
\label{sec:QDecays}

We assume, motivated by the 1-loop mass-shifts given in
subsection~\ref{masses}, that the spinless adjoints, $G_H^{(1,1)}$,
$W_H^{(1,1)}$, and $B_H^{(1,1)}$, are lighter than the $(1,1)$-quarks.
In this case, the KK quarks can decay into both vector and spinless
modes, via the KK-number preserving gauge interactions given in
\cite{Burdman:2005sr}.

The $SU(2)_W$-doublet (1,1) quarks can decay into a zero-mode quark
plus a $W^{(1,1)3}_{\mu}$ or $W^{(1,1)\pm}_{\mu}$ gauge boson, each
with a partial decay width given by
\bear
\Gamma\left(Q^{(1,1)}_+\!\to W^{(1,1) i}_{\mu} q_{L} \right) & \!=\! &
\frac{\alpha M_{Q_+^{(1,1)}} }{16 \sin^2\!\theta_{w}}
\left(1 - \frac{M_{W^{(1,1)}}^2}{M_{Q_+^{(1,1)}}^2} \right)^{\! 2} 
\nonumber \\ [0.5em] 
&& \times\,
\left(1 + \frac{M_{Q_+^{(1,1)}}^2}{2M_{W^{(1,1)}}^2}\right) ~,
\label{QtoWq}
\eear
where we neglected the final quark mass.  When the final state
includes the top quark, as for $Q_{L}^{3(1,1)}\!\to
W^{(1,1)i}_{\mu} t_{L}$ with $i=\pm,3$, this approximation may
not be valid.  In fact, for the 1-loop masses of subsection~\ref{masses},
these decays are closed for $1/R \sim 650~{\rm GeV}$, and are phase
space suppressed if the compactification scale is not much higher.  We
will neglect these decay channels in the following, although they
could be important at the LHC, where compactification scales well
above this limit can be probed.

Both the $SU(2)_W$-doublet and singlet (1,1) quarks may decay 
into $B^{(1,1)}_{\mu}$ and a zero-mode quark with a decay width of
\bear
\hspace{-1cm}
\Gamma\left(Q^{(1,1)}\!\to B^{(1,1)}_{\mu} q \right) & \!=\! &
\frac{y_{q}^2\alpha M_{Q^{(1,1)}} }{16 \cos^2\!\theta_{w}}
\left(1 - \frac{M_{B^{(1,1)}}^2}{M_{Q^{(1,1)}}^2} \right)^{\! 2} 
\nonumber \\ [0.5em] 
&& \hspace{.2cm} \times\,
\left(1 + \frac{M_{Q^{(1,1)}}^2}{2M_{B^{(1,1)}}^2}\right)~,
\label{QtoBq}
\eear
where $y_{q}$ is the hypercharge of the quark $q$ (normalized such that 
$y_{u_R} = 4/3)$.

\begin{table}[t]
\centering
\renewcommand{\arraystretch}{1.7}
\begin{tabular}{|c||c|c|c|c|}
\hline
\ decay modes \ & \ $Q_+^{(1,1)}$ \ & \ $B_+^{(1,1)}$ \ 
& \ $U_-^{(1,1)}$ \ & \ $D_-^{(1,1)}$ \ 
\rule{0mm}{5mm}\rule{0mm}{-22mm} \\ \hline 
     $G_H^{(1,1)} q$ & 41  & 58  & 61  & 86 \\ \hline
    $W_H^{(1,1)3} q$ & $\;8$ & 11  & --  & -- \\ \hline
  $W_H^{(1,1)\pm} q$ & 17  & 14 & --  & -- \\ \hline
     $B_H^{(1,1)} q$ & $\;\;\;\;0.3$ & $\;\;\;\;0.4$ & 13  & $\;5$ \\ \hline
  $W_\mu^{(1,1)3} q$ & 11  & 15  & --  & -- \\ \hline 
$W_\mu^{(1,1)\pm} q$ & 22  & --  & --  & -- \\ \hline 
   $B_\mu^{(1,1)} q$ & $\;\;\;\;0.9$ & $\;\;\;\;1.1$ & 26  & \ 9 \\ \hline
\end{tabular}
\medskip
\caption{Branching fractions (in percentage) into vector and spinless 
modes for the $SU(2)_{W}$-doublet quarks of the 
first two generations $Q_+^{(1,1)}$, for
the (1,1) mode of the $b_L$-quark, $B^{(1,1)}_+$, 
and for the $SU(2)_{W}$-singlet quarks, $U_-^{(1,1)}$
and $D_-^{(1,1)}$.  
The branching fractions of the $t$ quark (1,1) modes  are strongly 
dependent on $m_t R$, and are not shown here.
The phase-space suppression used for the
decay $B_+^{(1,1)}\!\rightarrow W^{(1,1)-}_{H}t_L$ 
corresponds to $1/R = 0.5$ TeV. 
}
\label{tab:BRQs}
\end{table}

Given that the quark (1,1) modes appear to be heavier than 
the spinless (1,1) gluon, the decay into a $G_H^{(1,1)}$
and a jet has a rather large width:
\be
\Gamma\left(Q^{(1,1)}\!\to G_H^{(1,1)} q\right) = \frac{\alpha_s}{6}\,
M_{Q^{(1,1)}} P\left(\frac{M_{G_H^{(1,1)}}}{M_{Q^{(1,1)}}},
\frac{m_q}{M_{Q^{(1,1)}}}\right) 
\label{QtoGHq}
\ee
where we defined the function
\be
P(x,y) = \left( 1-x^2 +y^2\right) \left[ \left( 1- x^2 -y^2 \right)^2
-4x^2y^2 \right]^{\!1/2} ~.
\ee
Note that the final-state quark is left-handed (right-handed) if the 
decaying (1,1) quark is an $SU(2)_W$-doublet (singlet).
The $SU(2)_W$-doublet (1,1) quarks may also decay into 
a $W_H^{(1,1)}$ and a jet,
\bear\hspace*{-4em}
\Gamma\left(Q_+^{(1,1)}\!\to W_H^{(1,1)i} q_{L}\right) &=& 
\frac{\alpha }{32\sin^{2}\!\theta_{w}}M_{Q_+^{(1,1)}}
\nonumber \\ [0.6em]
&& \hspace*{-4em}
\times\, P\left(\frac{M_{W_H^{(1,1)}}}{M_{Q_+^{(1,1)}}},
\frac{m_q}{M_{Q_+^{(1,1)}}}\right)  ~.
\label{QtoWHq}
\eear
We included here the dependence on the final-state quark mass,
$m_q$,  because the decay of the (1,1) $b_L$-quark, $B_+^{(1,1)}$,
is sensitive to $m_t R$. 
Any of the (1,1) quarks may also decay into the hypercharge spinless
adjoint and a jet with a width  
\bear \hspace*{-4em}
\Gamma\left(Q^{(1,1)}\!\to B_H^{(1,1)} q\right) &=& 
\frac{y^{2}_{q}\alpha }{32\cos^{2}\!\theta_{w}}} M_{Q^{(1,1)}
\nonumber \\ [0.6em]
&& \hspace*{-4em}
\times\, P\left(\frac{M_{B_H^{(1,1)}}}{M_{Q^{(1,1)}}},
\frac{m_q}{M_{Q^{(1,1)}}}\right) ~.
\label{QtoBHq}
\eear

\vspace*{-2mm}

\noindent
The decay widths
into the electroweak spinless adjoints are comparable to the 
weak decays of Eq.~(\ref{QtoWq}) or (\ref{QtoBq}),
while the decay into the spinless (1,1) gluon dominates. 
We base our estimates of the
$(1,1)$ quark branching fractions on the 1-loop corrected masses given in
subsection~\ref{masses}, and summarize them in Table~\ref{tab:BRQs}.


The decays into the spinless adjoints, $G_{H}^{(1,1)}$,
$W_{H}^{(1,1)3}$ and $B_{H}^{(1,1)}$, are very interesting since these
subsequently decay most of the time into a pair of top quarks, 
giving rise to a
potentially unique signal for these intrinsically 6D excitations.
This is due to the fact that the coupling of the spinless adjoints to
fermions is proportional to the fermion mass, as explained after
Eq.~(\ref{GH-coupling}).

As mentioned in Section II, the strongly interacting particles
receive important contributions from multiloop effects, and these
could in principle make the spinless gluons, $G_{H}^{(1,1)}$, heavier than the
$(1,1)$ quarks, thus closing these decay channels.  In that case, 
the $SU(2)_{W}$-doublet quarks would decay about $56\%$
of the time into $W_{\mu}^{(1,1)} q_{L}$, $42\%$ into $W_{H}^{(1,1)}
q_{L}$, and the rest into $B_{\mu}^{(1,1)} q_{L}$ and $B_{H}^{(1,1)}
q_{L}$.  The $SU(2)_{W}$-singlet quarks would decay about $67\%$ of the time
into $B_{\mu}^{(1,1)} q_{R}$ and $33\%$ into $B_{H}^{(1,1)} q_{R}$.

The $(1,1)$ leptons can decay to the (1,1) modes of the 
electroweak gauge bosons or spinless adjoints.  
The decay widths are given by Eqs.~(\ref{QtoBq}), (\ref{QtoWHq}) 
and (\ref{QtoBHq}), with $M_{Q^{(1,1)}}$ replaced by $M_{L^{(1,1)}}$,
 and $y_{q}$ replaced by $y_{l}$.  Using the
one-loop results for the various masses given in
subsection~\ref{masses}, we find the branching fractions summarized in
Table~\ref{tab:BRLs}.
\begin{table}[t]
\centering
\renewcommand{\arraystretch}{1.7}
\begin{tabular}{|c|c|c|c|c|}
\hline
 decay modes: & \ $W_H^{(1,1)\pm} l$ \ & \ $W_H^{(1,1)3} l$ \ 
& \ $B_H^{(1,1)} l$ \  & \ $B_\mu^{(1,1)} l$ \ \\ \hline\hline
$L^{(1,1)}$ & 45 & 22 & 21 & 12 \\ \hline
$E^{(1,1)}$ & --- & --- & 79 & 21 \\ \hline
\end{tabular}
\medskip
\caption{Branching fractions (in percentage) 
into vector and spinless modes for the
$SU(2)_{W}$-doublet leptons, $L^{(1,1)}$, and the $SU(2)_{W}$-singlet
leptons, $E^{(1,1)}$. }
\label{tab:BRLs}
\end{table}

Combining the $W_\mu^{(1,1)3}$ branching fractions into a (1,1) lepton
and a zero-mode lepton with the $L^{(1,1)}$ branching fractions,
leads to the cascade decays of $W_\mu^{(1,1)3}$
into (1,1) bosons shown in Table~\ref{tab:BRw}.
Given that $W^{(1,1)3}_{H}$
and $B^{(1,1)}_{H}$ decay almost exclusively into $t\bar{t}$ pairs,
about 21\% of the $W^{(1,1)3}_{\mu}$ decays lead eventually to
$t\bar{t}$ pairs.

\begin{table}[t]
\centering
\renewcommand{\arraystretch}{1.7}
\begin{tabular}{|c||c|c|}
\hline
\ decay modes \ &  \ $W_\mu^{(1,1)3}$ \ 
& \ $B_\mu^{(1,1)}$ \ 
\rule{0mm}{5mm}\rule{0mm}{-22mm} \\ \hline 
 $t\bar{t}$ & $13\;$ & $26 \;$\\ \hline
 $b\bar{b}$ & $16\;$ & 8\\ \hline
{\rm dijet} (no $b\bar{b}$) & $52\;$ & $62 \;$\\ \hline
 $\sum \ell^+\ell^-$ & $\;\;\;\;\;0.05$ & 3\\ \hline
 $\nu\bar{\nu}$ & $\;\;\;\;\;0.05$ & 1\\ \hline
    $W_H^{(1,1)3} + {\rm leptons}$ & 4 & --\\ \hline
  $W_H^{(1,1)\pm} + {\rm leptons}$ & $9$ & --\\ \hline
     $B_H^{(1,1)} + {\rm leptons}$ & $4$ & --\\ \hline
   $B_\mu^{(1,1)} + {\rm leptons}$ & $2$ & --\\ \hline
\end{tabular}
\medskip
\caption{Branching fractions of $W_\mu^{(1,1)3}$ and $B_\mu^{(1,1)}$
in percentage. The final states involving (1,1) bosons are
due to cascade decays via a (1,1) lepton.
The phase-space suppression of the decays into $t\bar{t}$ 
are computed for $1/R = 0.5$ TeV. 
}
\label{tab:BRw}
\end{table}
%

\subsection{Branching fractions of the gluon (1,1)-mode} 
\label{sec:GDecays}

The decays of $G_\mu^{(1,1)}$ that preserve KK numbers are into
$U_{-_R}^{(1,1)} \overline{u}_R$, $\overline{U}_{-_R}^{(1,1)} u_R$,
$U_{+_L}^{(1,1)} \overline{u}_L$, $\overline{U}_{+_L}^{(1,1)} u_L$,
and the analogous pairs of the other five quark flavors.
The partial widths for these decays are given by
\bear
\label{partial-width}
\Gamma\left(G_\mu^{(1,1)}\! \rightarrow U_{-_R}^{(1,1)} 
\overline{u}_R\right) & \!\simeq \!&
\frac{\alpha_{s}}{12} M_{G^{(1,1)}} 
\left(1-\frac{M_{U_-^{(1,1)}}^2}{M_{G^{(1,1)}}^2}\right)^{\!2} 
\nonumber \\ [0.5em]
&& \times\,\left(1+\frac{M_{U_-^{(1,1)}}^2}{2 M_{G^{(1,1)}}^2}\right) ~.
\eear
and analogous expressions for the other (1,1) quarks.
Given the approximate degeneracy of the gluon and quark KK modes, the
above decays are phase-space suppressed, and are in competition with
the decay into a quark-antiquark pair, which is suppressed by the
small KK-number violating couplings:
\bear
\label{dijet}
\Gamma\left(G_\mu^{(1,1)}\! \rightarrow q\overline{q} 
\right) & \!\simeq & 
\frac{\alpha_{s}}{12} \left(C^G\right)^2  
\left[ \left( \xi_{q_L}^G \right)^2 + \left( \xi_{q_R}^G \right)^2\right] 
 M_{G^{(1,1)}}
\nonumber \\ [0.5em]
&& \times \left(1 - \frac{3 m_q^2}{M_{G^{(1,1)}}^2} \right)
 ~,
\eear
where there is no flavor sum over the $q\overline{q}$ pairs.  The
decay into two gluons is also allowed \cite{Ponton:2005kx}, but even
further suppressed.

Using the parametrization for the (1,1)-mode
masses shown in Eqs.~(\ref{G-mass}) and (\ref{Q-mass}), 
assuming for the moment that all quark (1,1) modes are 
degenerate, and expanding for simplicity in $A_GC^G$, 
we find the total width of $G_\mu^{(1,1)}$:
\bear 
\Gamma_G &\!\approx\!& \alpha_s \left(C^G \right)^2  M_{G^{(1,1)}}
\left\{ \frac{1}{12}\left[ 4 \left(\xi_{q_L}^G \right)^2 
+ 2 \left(\xi_{u_R}^G \right)^2 \right. \right.
\nonumber \\ [0.5em]
&& \hspace{-1cm} \left. \left. \mbox{} 
+ 3 \left(\xi_{d_R}^G \right)^2  + 2\left(\xi_{t_L}^G \right)^2 
+ \left(\xi_{t_R}^G \right)^2  \right] 
+ 10 \left( A_G - A_Q\right)^2 \right\}
\nonumber \\
\eear
Here we have taken into account that the $G^{(1,1)}_\mu$ decays to 
a $t$ quark and one of its (1,1) modes
are kinematically forbidden for $1/R\sim 1$ TeV.
For $(A_G -A_Q)$ of order unity, $G_\mu^{(1,1)}$ has a width of the 
order of a few
percent of its mass.  Given that the $\xi_{q}^G$ coefficients are also
expected to be of order unity, the decay into a (1,1)-mode quark and a
zero-mode quark dominates.  For each flavor of $q\overline{q}$ pairs,
the branching fractions is approximately given by
\be
{\rm Br} \left( G_\mu^{(1,1)}\! \rightarrow q\overline{q} \right) 
\approx \frac{\left(\xi_{q_L}^G \right)^2 
+ \left(\xi_{q_R}^G \right)^2 }{120\left( A_G - A_Q\right)^2} ~.
\ee
%
%
which typically leads to branching fractions of less than $1\%$.  

\begin{table}[t]
\centering
\renewcommand{\arraystretch}{1.7}
\begin{tabular}{|c||c|}
\hline
\ decay modes \ & \ $G_\mu^{(1,1)}$ \
\rule{0mm}{5mm}\rule{0mm}{-22mm} \\ \hline 
    $G_H^{(1,1)} + {\rm jets}$ & 60.5\;\;  \\ \hline
    $W_H^{(1,1)3} + {\rm jets}$ & 3.2  \\ \hline
\ \ $W_H^{(1,1)\pm} + {\rm jets}$ \ \ & 6.1 \\ \hline
     $B_H^{(1,1)} + {\rm jets}$ & 4.8 \\ \hline
  $W_\mu^{(1,1)3} + {\rm jets}$ & 4.3   \\ \hline 
$W_\mu^{(1,1)\pm} + {\rm jets}$ & 7.0   \\ \hline 
   $B_\mu^{(1,1)} + {\rm jets}$ & 9.3 \\ \hline
 $t\bar{t}$ & $\; 0.5$  \\ \hline
 $b\bar{b}$ & $\; 0.8$  \\ \hline
{\rm dijet} (no $b\bar{b}$) & $3.3$  \\ \hline
\end{tabular}
\medskip
\caption{Branching fractions of $G_\mu^{(1,1)}$ in percentage. 
The final states involving (1,1) bosons are
due to cascade decays via a (1,1) quark.
With the exception of the decays into $W_{H,\mu}^{(1,1)\pm}$ and
$t\bar{t}$, whose widths are computed for $1/R = 0.5$ TeV,
the branching fractions are only mildly dependent on $1/R$.
}
\label{tab:BRg} 
\end{table}

The values of the $\xi$ parameters given by Eqs.~(\ref{xig-predict})
are indeed of order unity: $\xi_{q_{L}}^{G} = 0.74$,
$\xi_{u_{R}}^{G} = 0.95$, $\xi_{d_{R}}^{G} = 0.99$, $\xi_{t_{L}}^{G} =
0.61$ and $\xi_{t_{R}}^{G} = 0.69$.
However, according to the estimates of section~\ref{masses}, 
$A_G$ has a rather large value of $13/3$, so that 
for more precision we compute 
the branching fractions without expanding in $A_G C^G$.
Using the quark (1,1) masses given by Eqs.~(\ref{Q-mass}) and (\ref{aq}),
we find that the $G_\mu^{(1,1)}$ decays approximately 
$96\%$ of the time into a $(1,1)$ mode quark and the corresponding
 zero-mode quark. These decays are split as follows:
35\%  into down-type $SU(2)_W$-singlets, 
22\% into $u_R$ and $c_R$ modes, 
32\% into doublets of the first two generations,
and 6.3\% into $b_L$ modes.
The subsequent decay of the (1,1) quark depends on its transformation
properties under $SU(2)_W$, as discussed in
subsection~\ref{sec:QDecays}. Using the branching fractions of the
$(1,1)$ quarks given in Table~\ref{tab:BRQs}, we find 
the branching fractions for the $G_\mu^{(1,1)}$ cascade decays and
direct decays listed in Table~\ref{tab:BRg}.
The total width is $\Gamma_G \simeq 3.7\times 10^{-2} M_{G^{(1,1)}}$.

As mentioned in Section \ref{sec:QDecays}, 
the electrically-neutral spinless adjoints, $G^{(1,1)}_{H}$, $W^{(1,1)3}_{H}$
and $B^{(1,1)}_{H}$, decay most of the time into $t\bar{t}$ pairs.
The additional possible two-jet final states coming from two gluons are 
forbidden due to the vanishing of the operators similar to the one
shown in Eq.~(\ref{thisiszero}). 
Furthermore,  about 21\% of the $W^{(1,1)3}_{\mu}$ decays lead to 
$t\bar{t}$ pairs
(see section~\ref{sec:QDecays}),
while $B^{(1,1)}_{\mu}$ has a branching fractions of about 
$26\%$ into $t\bar{t}$.
The decays of $W^{(1,1)\pm}_{\mu}$ involving a (1,1) lepton
yield some additional $t\bar{t}$ pairs.
We therefore expect a significant fraction, of about
$72\%$, of the vector gluon modes to produce $t\bar{t}$ events. 

\section{Signals at the Tevatron}\label{sec:tevatron}
\setcounter{equation}{0}

In the absence of boundary terms, the conservation of KK number
implies that KK modes cannot be singly-produced.  In addition, as
pointed out in Ref.~\cite{Cheng:2002ab} for the 5D case, the nearly
degenerate spectrum typically results in rather soft jets and lepton
signals.  However, KK-number violating interactions such as those in
Eqs.~(\ref{bterms}) and (\ref{gluon-terms}), while still preserving
$Z_2^{\rm KK}$, allow for the production of single $(1,1)$ states
through their interactions with zero modes.  In what follows, we study
the $s$-channel production of the $(1,1)$ KK gluon $G^{(1,1)}_\mu$, as
well as of the electroweak gauge bosons $W^{(1,1)3}_\mu$ and
$B^{(1,1)}_\mu$.  Their subsequent decays give rise to interesting
signals at the Tevatron.

\subsection{$s$-channel production of the (1,1) modes}

Let us first consider the $s$-channel production of the gluon vector
mode $G_\mu^{(1,1)}$ through the coupling to $q\bar{q}$ pairs given in
Eq.~(\ref{coupling}).  The differential cross section for the
$s$-channel process $q\bar{q} \rightarrow G_\mu^{(1,1)} \rightarrow
U_{-_R}^{(1,1)} \overline{u}_R$
is given by
\bear 
\frac{d\hat{\sigma}_G}{d(\cos\theta)}
& = & \frac{\pi\alpha_s^2}{36\hat{s}^2}\! \left( C^G\right)^2 
\frac{\left(\hat{s}-M_{Q^{(1,1)}}^2\right)^{\!2}}
{\left(\hat{s}-M_{G^{(1,1)}}^2\right)^2 + M_{G^{(1,1)}}^2 \Gamma_G^2}
\nonumber \\ [0.5em]
& \times & \left\{ \rule{0mm}{3.5mm}  
\left[\hat{s}\left(1 - \cos\theta \right)^2
+ M_{Q^{(1,1)}}^2 \sin^2\!\theta \right] \left( \xi_{q_L}^G \right)^2\right.
\nonumber \\[0.5em]
& + & \left. \rule{0mm}{3.5mm}  
\left[\hat{s}\left(1 +\cos\theta \right)^2
+ M_{Q^{(1,1)}}^2 \sin^2\!\theta \right] \left( \xi_{q_R}^G \right)^2 
\right\}~.
\nonumber\\
\label{parton}
\eear
where $\theta$ is the angle between the momenta of $U_{-_R}^{(1,1)}$
and $q$, and $\hat{s}$ is the energy of the parton collision, both
defined in the center of mass frame.  

In the narrow width approximation, the parton-level cross section for
the production of a (1,1) gluon takes a simple form:
\bear 
 \hat{\sigma}\!\left(q\bar{q} \rightarrow G_\mu^{(1,1)} \right)
& = & \frac{4\pi^2\alpha_s}{9 M_G} \left(C^G\right)^2  
\left[ \left(\xi_{q_L}^G\right)^2 + \left(\xi_{q_R}^G\right)^2 \right]
\nonumber \\[0.5em]
&& \times \delta\! \left(\sqrt{\hat{s}} - M_{G^{(1,1)}} \right) ~.
\label{parton-total}
\eear
Integrating this partonic cross section 
over the parton distribution functions, we find the inclusive cross
section. 
At the Tevatron, the total production cross-section is given by
\bear
\sigma\!\left(p\bar{p} \rightarrow G_\mu^{(1,1)} X \right)
& = &\frac{8\pi^2\alpha_s}{9 s} \left(C^G\right)^2 
\sum_q t_q\!\left(M_{G^{(1,1)}}^2/s\right)
\nonumber \\[0.5em]
&& \times
\left[ \left(\xi_{q_L}^G\right)^2 + \left(\xi_{q_R}^G\right)^2 \right] ~.
\label{g11xs}
\eear
To leading order in $\alpha_s$, 
\be
t_q(z) = \int_{z}^1 \frac{dx}{x}\left[ \rule{0mm}{3.5mm}  
q(x)\,q\left(z/x\right) + \overline{q}(x) \,
\overline{q}\left(z/x\right) 
\rule{0mm}{3.5mm}  \right] ~.
\ee
The parton distribution functions (PDF's)
$q(x)$ and $\overline{q}(x)$ are
evaluated at the scale $M_{G^{(1,1)}}/2$, and $\sqrt{s} = 1.96$ TeV in Run II.
We use the CTEQ6 leading order PDF's  \cite{Pumplin:2002vw}, 
and a correction factor of $K=1.3$ to approximate the next-to-leading order
(NLO) QCD corrections. This approximation is often used in 
the case of $Z^\prime$ production (for a discussion of its accuracy,
see Ref.~\cite{Carena:2004xs}). 
Note that $W^{(1,1)3}_\mu$ and $B^{(1,1)}_\mu$
fall into this category, whereas $G^{(1,1)}_\mu$ production 
has different color flow, so that a slightly different
$K$ factor may be necessary in that case; we will not
study this issue in what follows.
The result is the solid line shown in
Figure~\ref{g11sc}.

We emphasize that this is only a rough estimate of the vector mode
production cross sections. We have not included
several corrections: i) the non-resonant process induced by a
$t$-channel exchange of a (1,1) gluon which involves a single
KK-number violating interaction; ii) $s$-channel production of a (1,1)
gluon from gluon fusion, via dimension-6 operators 
(note that the $q\bar{q}$ initial state
dominates at the Tevatron);  
iii) exact NLO and next-to-next-to-leading order QCD corrections.  
However, we expect our
estimate to be correct up to a factor of less than 2, which is
sufficient for the purpose of deciding whether a search for (1,1)
modes at the Tevatron is useful.

\begin{figure}[t]
\hspace*{-.5em}
\psfrag{sigma}[b]{\hspace*{1.5em} $\sigma 
\left(\rule{0em}{0.5em}\right. p\overline{p}\rightarrow 
V^{(1,1)}_\mu \!\left.\rule{0em}{0.5em}\right)$ \ [fb]}
\psfrag{MR}[t]{\hspace*{.9em} $M_{V^{(1,1)}}$ [GeV]}
\psfig{file=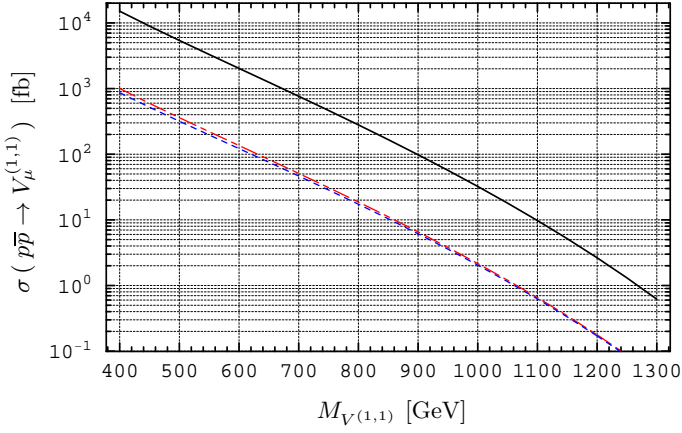,width=8.9cm,angle=0}
\vskip0.1cm
\caption{Production cross sections for $(1,1)$ vector modes in the 
$s$ channel at the Tevatron, as a function of their mass. 
The solid line is for $G^{(1,1)}_\mu$, while 
the dashed and dotted (lowest) lines are
for $W^{(1,1)3}_\mu$ and  $B^{(1,1)}_\mu$, 
respectively (accidentally, the cross sections for these two are 
close to each other such that they might not be distinguishable).}
\label{g11sc}
\end{figure}


The production cross sections for the $s$-channel processes $q\bar{q}
\rightarrow W_\mu^{(1,1)3}, B_\mu^{(1,1)}\!\!\rightarrow
q^\prime\bar{q}^\prime$ may be computed in a similar fashion.  The
differential cross sections for these two processes are given by
\bear \hspace*{-.4em}
\frac{d\hat{\sigma}_W}{d(\cos\theta)}
& = & \frac{\pi\alpha^2\left(C^G\right)^4 }{128\sin^4\!\theta_w} \,
\left(\xi_q^W \xi_{q^\prime}^W \right)^2  f_{q^\prime}\!\left(\cos\theta \right)
\nonumber \\ [0.5em]
&& \times  
\frac{\hat{s}}
{\left(\hat{s}-M_{W^{(1,1)}}^2\right)^2 + M_{W^{(1,1)}}^2 \Gamma_W^2} 
~,
\label{parton-W}
\eear
\bear 
\frac{d\hat{\sigma}_B}{d(\cos\theta)}
& = & \frac{\pi\alpha^2\left(C^G\right)^4 }{128\cos^4\!\theta_w} \;
\frac{\hat{s}}
{\left(\hat{s}-M_{B^{(1,1)}}^2\right)^2 + M_{B^{(1,1)}}^2 \Gamma_B^2} 
\nonumber \\ [0.5em]
&& \times \left[ \rule{0mm}{3.5mm}  
\left( a_{q_L}a_{q_L^\prime} + a_{q_R}a_{q_R^\prime}\right) 
f_{q^\prime}\!\left(\cos\theta\right) \right.
\nonumber \\ [0.5em]
&& \; + \left. \rule{0mm}{3.5mm}
\left( a_{q_R}a_{q_L^\prime} + a_{q_L}a_{q_R^\prime}\right) 
f_{q^\prime}\!\left(-\cos\theta\right) \right]
\label{parton-B}
\eear
where the function that encodes the angular distribution 
has the following form:
\be
f_q(y) = \left( 1 + y\right)^2 
- 2\left( 1 + 4 y + 3 y^2\right) 
\frac{m_{q}^2}{\hat{s}}
+ O\left( m_{q}^4/\hat{s}^2 \right) ~.
\ee
Note that we keep the dependence on the final-state quark masses,
which is useful for the decay into $t\overline{t}$.
The parameters $a_{q_L}$, $a_{q_L^\prime}$, $a_{q_R}$, $a_{q_R^\prime}$ 
are products of hypercharges and $\xi$ parameters:
\be
a_{q} = 
\left(\xi_{q}^B y_{q}\right)^{\!2} ~.
\ee

The parton-level production cross sections are
given in the narrow width approximation by 
\be 
\hat{\sigma}\!\left(q\bar{q} \rightarrow W_\mu^{(1,1)3} \right)
= \frac{\pi^2\alpha \left(\xi_q^W C^G\right)^2 }
{12 \sin^2\!\theta_w \, M_{W^{(1,1)}} }
 \, \delta\! \left( \sqrt{\hat{s}} - M_{W^{(1,1)}} \right)
\label{parton-total-w}
\ee
\bear 
\hat{\sigma}\!\left(q\bar{q} \rightarrow B_\mu^{(1,1)} \right)
&=& \frac{\pi^2\alpha \left(C^G\right)^2 }
{12 \cos^2\!\theta_w \, M_{B^{(1,1)}} }
\left(a_{q_L} + a_{q_R}\right) 
\nonumber \\ [0.5em]
&& \times \, \delta\! \left( 
\sqrt{\hat{s}} - M_{B^{(1,1)}} \right) ~.
\label{parton-total-b}
\eear

The total production cross-section at the Tevatron are given by
\bear
 \hspace*{-.1em} \sigma\!\left(p\bar{p} \rightarrow W_\mu^{(1,1)3} 
\right)
& = & \frac{\pi^2\alpha\left(C^G\right)^2 }{6\sin^2\!\theta_w \, s} 
\sum_q \left(\xi_q^W\right)^2 
\nonumber \\ [0.5em]
&& \times \,
t_q\!\left(
M_{W_\mu^{(1,1)}}^2/s\right)
\label{w11xs}
\nonumber \\ [1em]
 \hspace*{-.1em} 
\sigma\!\left(p\bar{p} \rightarrow B_\mu^{(1,1)} \right)
&  = & \frac{\pi^2\alpha\left(C^G\right)^2 }{6\cos^2\!\theta_w \, s} 
\sum_q \left(a_{q_L} + a_{q_R}\right) 
\nonumber \\ [0.5em]
&& \times \,t_q\!\left(M_{B_\mu^{(1,1)}}^2/s\right) ~,
\label{b11xs}
\eear
and are shown in Figure~{\ref{g11sc}.
Note that the $B_\mu^{(1,1)}$ production is suppressed compared to 
$W_\mu^{(1,1)3}$ production by a $\tan^2\!\theta_w$ factor, but it is 
also enhanced by the larger values of the $\xi$ parameters, such that 
the curves representing the two cross sections are very 
close to each other.

\subsection{Peaks in the invariant mass distributions}
\label{sec:limit}

\begin{figure}[t]
\begin{center}
\scalebox{0.65}{
\begin{picture}(100,120)(70,-30)
\thicklines
\put(0,0){\line(-1,1){50}}
\put(0,0){\line(-1,-1){50}}\put(0,0){\circle*{9}}

\multiput(0,0)(8, 0){9}{\qbezier(0,0)(5,5)(10,0)
                 \qbezier(10,0)(12,-5)(9,-7)\qbezier(9,-7)(6,-5)(8,0)}
\qbezier(72,0)(77,5)(81,0)
\put(80,0){\line(2,1){60}}
\put(140,30){\line(1,1){40}}
\multiput(140,30)(15,0){5}{\line(1,0){10}}
\put(210,30){\line(5,2){60}}
\put(210,30){\line(5,-2){60}}
\put(211,29){\circle*{8}}
\put(-57,34){\LARGE $q$}
\put(-57,-39){\LARGE $\overline{q}$}
\put(28,11){\LARGE $G_\mu^{(1,1)}$}
\put(89,22){\LARGE $U_-^{(1,1)}$}
\put(105,-41){\LARGE jet}
\put(150,67){\LARGE jet}
\put(160,6){\LARGE $G^{(1,1)}_H$}
\put(279,48){\LARGE $t$}
\put(279,-5){\LARGE $\overline{t}$}
\put(80,0){\line(2,-1){60}}
\end{picture}
}
\end{center}
\caption{Production of the vector (1,1) gluon followed by a cascade
decay.  The $\bullet$ stands for a KK number-violating coupling.
Other diagrams having the same topology exist: 
the $U_-^{(1,1)}$ quark KK mode may be replaced by $D_-^{(1,1)}$,
$Q_+^{(1,1)}$, or the corresponding anti-quarks; in addition
the spinless gluon $G^{(1,1)}_H$ may be replaced by 
$B^{(1,1)}_\mu$ or $B^{(1,1)}_H$, and in the case where 
the quark KK mode is  an $SU(2)_W$ doublet, by 
$W^{(1,1)3}_\mu$ or $W^{(1,1)3}_H$.}
\label{cascadefig}
\end{figure}
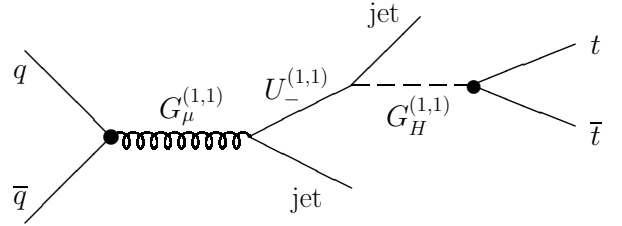

Once produced at the Tevatron, the $G^{(1,1)}_\mu$, $W^{(1,1)3}_\mu$,
and $B^{(1,1)}_\mu$ would decay with the branching fractions 
given in Tables~\ref{tab:BRg} and \ref{tab:BRw}. 
These vector (1,1) modes are leptophobic
(only $B^{(1,1)}_\mu$ has a potentially interesting 
branching fractions of about 1\% into each lepton pair), 
but have rather large branching fractions into $t\bar{t}$ pairs,
either directly or via cascade decays as explained 
at the end of Section~\ref{sec:GDecays}.
Altogether there are six resonances that can be produced 
in the $t\bar{t}$ invariant mass distribution: 
the vector and spinless (1,1) modes of the gluon and of the
two electroweak gauge bosons.
However, the decay $G^{(1,1)}_\mu\!\!\to\! t\bar{t}$ has a 
negligible branching fraction. 
Therefore, we will 
concentrate on the $t\bar{t}$ peaks at the  masses of 
$G^{(1,1)}_H$, $W^{(1,1)3}_\mu$, $B^{(1,1)}_\mu$, $W^{(1,1)3}_H$ and 
$B_H^{(1,1)}$.
These are given by
$1.10 M_{1,1}$, $1.08 M_{1,1}$, $0.98 M_{1,1}$, $0.95 M_{1,1}$ 
and $0.86 M_{1,1}$, 
where $M_{1,1} =\sqrt{2}/R$.  

\begin{figure}[t]
\begin{center}
\scalebox{0.65}{
\begin{picture}(150,210)(40,-10)
\thicklines
\put(-2,140){\line(-1,1){50}}
\put(-2,140){\line(-1,-1){50}}
\put(0,140){\circle*{9}}
\put(123,140){\circle*{9}}
\multiput(0,140)(20,0){6}{\qbezier(0,0)(5,-5)(10,0)\qbezier(10,0)(15,5)(20,0)}
\put(123,140){\line(2,1){60}}
\put(-57,174){\LARGE $q$}
\put(-57,101){\LARGE $\overline{q}$}
\put(18,151){\LARGE $W_\mu^{(1,1)3}$, $B_\mu^{(1,1)}$}
\put(189,162){\LARGE $t$}
\put(189,101){\LARGE $\overline{t}$}
\put(123,140){\line(2,-1){60}}
%
\put(0,0){\line(-1,1){50}}
\put(0,0){\line(-1,-1){50}}\put(0,0){\circle*{9}}
\multiput(0,0)(20,0){4}{\qbezier(0,0)(5,-5)(10,0)\qbezier(10,0)(15,5)(20,0)}
\put(80,0){\line(2,1){60}}
\put(140,30){\line(1,1){40}}
\multiput(140,30)(15,0){5}{\line(1,0){10}}
\put(214,30){\line(5,2){60}}
\put(214,30){\line(5,-2){60}}
\put(216,29){\circle*{8}}
\put(-57,34){\LARGE $q$}
\put(-57,-39){\LARGE $\overline{q}$}
\put(21,11){\LARGE $W_\mu^{(1,1)3}$}
\put(89,22){\LARGE $L_-^{(1,1)}$}
\put(105,-41){\LARGE $\ell$}
\put(154,60){\LARGE $\ell$}
\put(155,6){\LARGE $W^{(1,1)3}_H$}
\put(283,48){\LARGE $t$}
\put(283,-5){\LARGE $\overline{t}$}
\put(80,0){\line(2,-1){60}}
\end{picture}
}
\end{center}
\vspace*{2em}
\caption{Production of $W^3_\mu$ and $B_\mu$ (1,1) modes, 
followed by representative decays.}
\label{cascadefig-ew}
\end{figure}
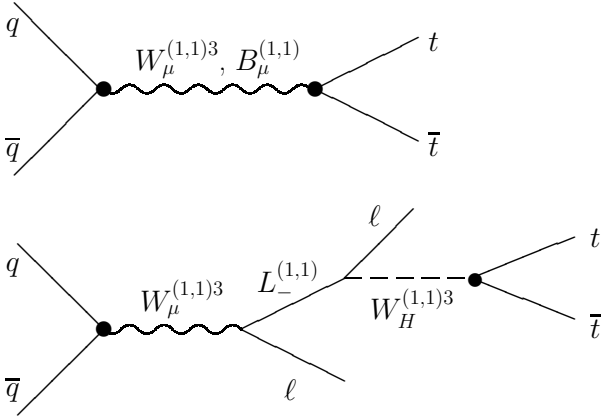

The spinless (1,1) gluon, $G^{(1,1)}_H$, is produced only in cascade decays 
of the vector (1,1) gluon, as shown in Figure \ref{cascadefig},
the electroweak spinless adjoints are produced 
in the cascade decays of both $W^{(1,1)3}_\mu$ (see Figure \ref{cascadefig-ew})
and $G^{(1,1)}_\mu$, while the electroweak vector modes are produced 
both in cascade decays and directly, as shown in Figure \ref{cascadefig-ew}.
The cross sections for producing $t\bar{t}$ pairs with
an invariant mass corresponding to the five resonances are given by
\bear
&& \sigma_{t\bar{t}}\left(G_H^{(1,1)}\right) = 
 \sigma\!\left(G_\mu\right)\,b_G\!(G_H) ~,
\nonumber \\ [0.6em]
&& \sigma_{t\bar{t}}\left(W_\mu^{(1,1)3}\right) = 
\left[ \sigma\!\left(G_\mu\right) \, b_G\!\left(W_\mu^3\right)\,
+ \sigma\!\left(W_\mu^{3}\right) \right] \,b_W\!\left(t\bar t\right)~,
\nonumber 
\eear
\bear
&& \sigma_{t\bar{t}}\left(B_\mu^{(1,1)}\right) =
\left[ \sigma\!\left(G_\mu\right) \, b_G\!\left(B_\mu\right)\,
 + \,\sigma\!\left(W_\mu^{3}\right) \, b_W\!\left(B_\mu\right) 
\nonumber \right. \\ [0.5em]
&& \hspace*{6em}\left. + \sigma\!\left(B_\mu\right) \,\right] \, b_B\!\left(t\bar t\right)\, ~,
\nonumber \\ [0.6em]
&& \sigma_{t\bar{t}}\left(W_H^{(1,1)3}\right) = 
\sigma\!\left(G_\mu\right) \, b_G\!\left(W_H^3\right)\,
+ \sigma\!\left(W_\mu^{3}\right) \, b_W\!\left(W_H^3\right)~, \nonumber\\ [0.6em]
&& \sigma_{t\bar{t}}\left(B_H^{(1,1)}\right) =
\sigma\!\left(G_\mu\right)\,b_G\!\left(B_H\right)
+ \sigma\!\left(W_\mu^{3}\right) \, b_W\!\left(B_H\right) ~,\nonumber\\
\label{peaks}
\eear
where we introduced the short-hand notations
\be
\sigma\!\left(V_\mu\right)
\equiv \sigma\!\left(p\bar{p} \rightarrow V_\mu^{(1,1)} X \right) ~,
\ee
for the production cross-sections shown in Figure~\ref{g11sc},
and 
\bear
&& b_G\!(V) \equiv 
{\rm Br}\left(G_\mu^{(1,1)}\!\to V^{(1,1)} +{\rm jets}\right) ~,
\nonumber\\ [0.5em]
&& b_W\!(V) \equiv
{\rm Br}\left(W_\mu^{(1,1)3}\!\to V^{(1,1)} +{\rm leptons}\right) ~,
\nonumber\\ [0.5em]
&& b_V\!\left(t\bar{t}\right) \equiv
{\rm Br}\left(V_\mu^{(1,1)}\!\to t\bar{t}\right) ~.
\eear
for the branching fractions given in Tables~\ref{tab:BRg} and \ref{tab:BRw}. 
In Eqn.~(\ref{peaks}) we have used branching fractions of 100\%
for electrically-neutral spinless adjoints
into $t\overline{t}$, which is a reasonably good approximation.

Additional contributions to the $t\overline{t}$ peaks at 
the $B_\mu^{(1,1)}$, $W_H^{(1,1)3}$ and $B_H^{(1,1)}$ masses
come from $s$-channel production of $W_\mu^{(1,1)\pm}$ 
followed by cascade decays similar to the one in Figure~\ref{cascadefig-ew}.
However, the relevant branching fractions for $W_\mu^{(1,1)\pm}$ are 
at most a few percent, and for simplicity we ignore them.
We have also neglected contributions to Eq.~(\ref{peaks}) coming from the 
cascade decays of a $(1,1)$ KK gluon through a $W_\mu^{(1,1)}$
into $B_\mu^{(1,1)}$, $W_H^{(1,1)3}$ or $B_H^{(1,1)}$, because these are  
suppressed by an additional branching ratio. 

The five resonances described above are very narrow, but 
cannot be separately resolved at hadron collider experiments. 
At CDF and D0, the $t\bar t$ pair mass
resolution is expected to be around $10\%$, so one could hope for 
at most three distinct peaks. The heaviest one
corresponds to the $G_H^{(1,1)}$ and $W_\mu^{(1,1)3}$ resonances
which have masses 2\% apart, with an average of $1.09 M_{1,1}$.
Then, there is a peak at $0.97 M_{1,1}$, 
composed of $W_H^{(1,1)}$ and $B_\mu^{(1,1)}$, 
whose masses separated by 3\% cannot be resolved experimentally. 
The third peak, due to $B_H^{(1,1)}$, is at $0.86M_{1,1}$.

\begin{figure}
\psfrag{sigma}[B]{\hspace*{2em} $\sigma
\left(\rule{0em}{0.5em}\right. p\overline{p}\rightarrow V^{(1,1)}
 \rightarrow t\overline{t} \left.\rule{0em}{0.5em}\right) $ \ [fb]}
\psfrag{Mtt}[t]{$M\!\left(t\overline{t}\right)$ \ [GeV]}
\hspace*{-0.9em}
\psfig{file=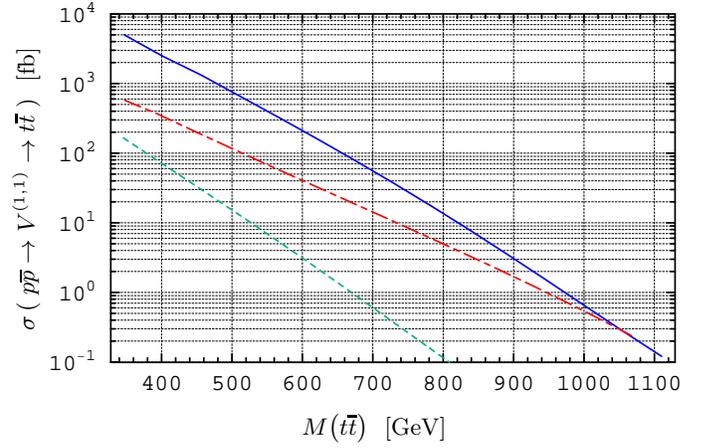,width=9.1cm,angle=0}
\vspace*{-1mm}
\caption{Cross section for the production of $t\bar t$ pairs
at the Tevatron from the three distinct mass peaks: 
$G_H^{(1,1)}+W_\mu^{(1,1)3}$ (top, solid line),
$W_H^{(1,1)3}+B_\mu^{(1,1)}$ (middle line) and $B_H^{(1,1)}$ 
(bottom line).}
\label{x112q}
\end{figure}

In Figure~\ref{x112q} we plot the cross sections for $t\bar t$ pairs
coming from the three mass peaks. 
The current preliminary limits at the 95\% confidence level 
from D0 \cite{d0} and CDF \cite{cdf} on the production 
cross section of a narrow $t\bar t$ resonance,
based on 0.37 fb$^{-1}$ and 0.32 fb$^{-1}$ of Run II data, respectively, 
are around 1 pb for $M\!\left(t\overline{t}\right)$ above 600 GeV or so,
and a few times larger than that for $M\!\left(t\overline{t}\right)$
in the 350 -- 550 GeV range due to some excess events.
The $G_H^{(1,1)}+W_\mu^{(1,1)3}$ and
$W_H^{(1,1)3}+B_\mu^{(1,1)}$ mass peaks have cross sections 
not far below these limits, but at the moment $1/R$ is not constrained by 
$s$-channel production of (1,1) modes.

Nevertheless, the much larger integrated luminosity expected until the end
of Run II will make it possible to probe an interesting range of values for
the compactification scale.
In order to estimate the ultimate reach of the Tevatron, 
we plot in Figure~\ref{sum} the sum of the cross sections into 
$t\bar{t}$ pairs from all peaks versus the uncorrected mass 
of the $(1,1)$ KK level, i.e., $M_{1,1}=\sqrt{2}/R$.  
One should keep in mind that for any
given value of $1/R$ the separation between consecutive mass peaks is
slightly above 10\%. For instance, for $1/R=500~$GeV, the peaks are at 
770 GeV, 680 GeV, and 610 GeV.
For this value of the compactification scale, 
the total cross section for $t\bar t$ pairs from (1,1) resonances
is approximately 40~fb. 
The branching fraction for $t\overline{t}$ into the 
lepton plus jets final state used in \cite{d0,cdf} is 29\%, 
while the product of 
acceptance times efficiency is expected to be in the 15\%-20\% range.
Therefore, approximately 5\% of the $t\overline{t}$ pairs 
can be selected, so that 
an integrated luminosity of 5 fb$^{-1}$ will 
result in a total of about 10
reconstructed $t\bar t$ events from the sum of all (1,1) resonances,
for $1/R=500~$GeV.

Although the background is small, due especially to standard model
$t\overline{t}$ and $W + 4j$ productions \cite{d0}, 
it is not negligible at large luminosity 
for $M\!(t\overline{t})$ below 800 GeV or so, and therefore 
the ultimate Tevatron reach is likely to be below $1/R=500~$GeV.

An estimate of the future sensitivity to $t\overline{t}$ resonances 
\cite{rossin} shows that with 4 fb$^{-1}$ the production cross section 
will be down to 1.3 pb for a mass of 450 GeV, and 0.7 pb for 
a mass of 550 GeV. Comparing these numbers with the cross section for the 
$G_H^{(1,1)}+W_\mu^{(1,1)3}$ mass peak given in Figure 6,
we find that there will be sensitivity to a peak of mass up to 480 GeV.
Given that Run II may deliver more than 4 fb$^{-1}$, and that 
there are additional $t\overline{t}$ events from the nearby 
$W_H^{(1,1)3}+B_\mu^{(1,1)}$ mass peak, it is likely that 
the ultimate Tevatron sensitivity will be for a $G_H^{(1,1)}+W_\mu^{(1,1)3}$ 
mass peak above 500 GeV, corresponding to a limit of 320 GeV on $1/R$.

A cautionary comment needs to be made:
the preliminary D0 and CDF limits mentioned above have been derived 
based on the assumption that there is a single $t\overline{t}$
resonance having a width equal to 1.2\% of its mass \cite{d0, cdf}.
In the model with two universal extra dimensions discussed here,
there are several resonances arising from both direct production
and cascade decays, and therefore one would need to set
limits based on these facts. However, the extra jets 
and leptons that are produced in the cascade decays are relatively 
soft due to the approximate mass degeneracy of the (1,1)
modes, and are not likely to change dramatically the limits.
The individual resonances are very narrow (with 
widths of at most 0.1\% of their mass for the electroweak
KK bosons, and of the order of 1\% for $G_H^{(1,1)}$),
but the main ones come in pairs, with separations of 2--3\%
within the pairs. Given the expected 
resolution of 10\%, such a pair of resonances looks like a single 
resonance, similar to the one used to set the CDF and D0 limits. 
The presence of two pairs (the top two curves in Figure 6)
with comparable cross sections, which may partially overlap,
is likely to have a stronger impact on the limits. 
Overall though this is not a concern at the level of accuracy 
employed here. 

\begin{figure}
\psfrag{sigma}[B]{ \parbox[t]{10cm}{.\\ [-.9em] \hspace*{1.5em} $\sigma
\left(\rule{0em}{0.5em}\right. p\overline{p} \rightarrow 
\sum  V^{(1,1)}\rightarrow t\overline{t}
\left.\rule{0em}{0.5em}\right)$ \ [fb]} }
\psfrag{M11}[t]{
\hspace*{1.5em}{\large $\frac{\sqrt{2} \vphantom{\sqrt{F_F}} }{R}$} \ [GeV]}
\hspace*{-0.8em}
\psfig{file=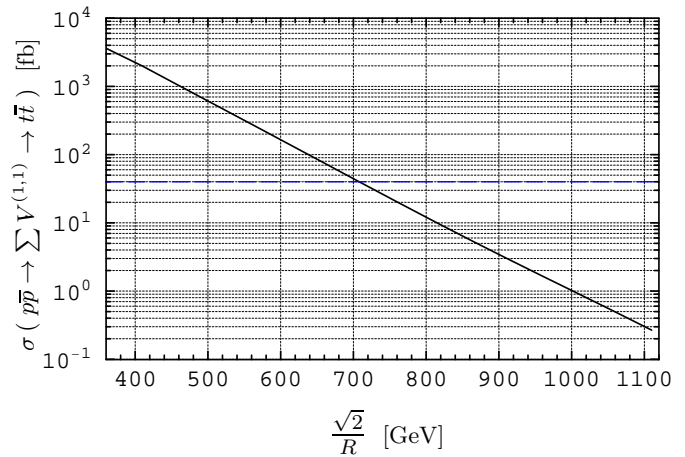,width=8.9cm,angle=0}
\vspace*{1mm}
\caption{Cross section for the production of $t\bar t$ pairs
at the Tevatron 
from the sum of all mass peaks at KK level (1,1),
as a function of $M_{1,1}=\sqrt{2}/R$. 
The dashed horizontal line marks 10 selected events with $5 {\rm fb}^{-1}$,
assuming that 5\% of the $t\overline{t}$ pairs are selected.
}
\label{sum}
\end{figure}

So far, we have assumed in this Section that the 
KK mass splittings and KK-number violating 
couplings are given by the one-loop effects discussed
in Section~\ref{sec:6DSM}.
We reiterate that there are uncertainties in the mass splittings
and couplings to zero modes of the (1,1) modes due to higher loops
involving the $SU(3)_C$ interactions (see Section~\ref{sec:local}).
These could have several effects. For example, 
the mass of $G_H^{(1,1)}$ could be further apart from the mass
of $W_\mu^{(1,1)3}$. However, it turns out that 
the majority of events in the 
$G_H^{(1,1)}+W_\mu^{(1,1)3}$ mass peak are due to $G_H^{(1,1)}$,
so that the uncertainty in the mass of $G_H^{(1,1)}$ does 
not result into a changed sensitivity to the highest peak, 
but rather into an uncertainty on the limit on $1/R$.
A more drastic effect of the higher loops would be to invert the mass
hierarchy between the (1,1) quarks and $G_H^{(1,1)}$. In that case
the $t\overline{t}$ peak at the $G_H^{(1,1)}$ mass would be highly
suppressed, but a large fraction of the $G_\mu^{(1,1)}$ cascade decays 
would result in $t\overline{t}$ events at the mass peaks due to the
(1,1) electroweak bosons (see Section~\ref{sec:QDecays}).

If the qualitative mass hierarchy of (1,1) modes is the one given 
by the one-loop results, then shifts in two parameters 
from higher loops could have a substantial impact on the production cross
section of the $t\bar t$ resonances. First, the mass splitting between
$G_\mu^{(1,1)}$ and $G_H^{(1,1)}$ could be different. If it is 
larger (smaller), then the cross section for a given $G_H^{(1,1)}$ 
mass decreases (increases), as it is harder (easier) to produce the 
$G_\mu^{(1,1)}$ boson,
which is the main source of $G_H^{(1,1)}$ production, via cascade decays.

Second, the average coupling of $G_\mu^{(1,1)}$ to quark zero-modes,
$\xi_q^G$ [see Eqs.~(\ref{xi-def}) and (\ref{coupling})],
may also be different than the one-loop result. 
Assuming that the change in $\xi_q^G$ is at most a factor of 2, 
the cross section for the $t\bar t$ signal due to 
$G_H^{(1,1)}$ decays (approximately given by the top curve
in Figure 6) could still increase by a factor of 4, because the 
cross section is proportional to $(\xi_q^G)^2$, as shown in 
Eq.~(\ref{g11xs}). In that case the CDF and D0 sensitivity from 
Ref.~\cite{rossin} with 4 fb$^{-1}$ would increase to 
a $G_H^{(1,1)}$ mass of about 700 GeV. 
Notice also that a large decrease in $\xi_q^G$ would not necessarily 
dilute completely the Tevatron reach, since the 
$W_\mu^{(1,1)}$ and $B_\mu^{(1,1)}$ production cross sections 
depend on parameters other than $\xi_q^G$ (those parameters, 
$\xi_q^W$ and $\xi_q^B$,
are affected by higher loops that change the wave 
function renormalization of the quarks, but their shifts may be 
different than for $\xi_q^G$).
Based on these considerations, we conclude that the ultimate
Tevatron sensitivity to $t\bar t$ mass peaks from the 6D Standard
Model may be as high as in the 0.5--0.7 TeV range (corresponding to
$1/R$ as high as 320--450 GeV), depending on the values of parameters 
controlled by the $SU(3)_C$ interactions of the KK modes.

For low enough $1/R$ one may hope for a discovery of  $t\bar t$ resonances
at the Tevatron. In that case one should find 
ways of discriminating the models with two universal extra dimensions
against other models that predict $t\overline{t}$ resonances,
such as Topcolor \cite{Hill:1991at} or certain technicolor models 
\cite{Lane:1989ej}.
Fortunately, the extra jets and leptons from cascade decays
may provide useful checks 
for confirming that the resonances are due to (1,1) modes.
The jets come from decays of the colored (1,1) states 
as shown in Figure 4, and may carry an energy of up to $10-15$\% of the 
mass of the decaying particle. The leptons come from cascade 
decays of $W_\mu^{(1,1)3}$ (see Figure 5), with rather small
but still relevant 
branching fractions given in Table \ref{tab:BRw}. 
Measurements of angular distributions may further discriminate among 
various models.

In addition to the decays into $t\bar t$ pairs from the above
mentioned resonances, there will be decays of $W_\mu^{(1,1)}$, 
$B_\mu^{(1,1)}$, and 
(to a lesser extent, see Table~\ref{tab:BRg}) 
of $G_\mu^{(1,1)}$ into a pair of
jets. From Table~\ref{tab:BRw}, we see that 
Br$(W_\mu^{(1,1)3}\! \to\! {\rm dijets})= 64\%$ and 
Br$(B_\mu^{(1,1)}\! \to\! {\rm dijets})= 71\%$, where we included $b$ jets.
Figure~\ref{g11sc} shows that dijet resonances
at the $W_\mu^{(1,1)3}$ and $B_\mu^{(1,1)}$ masses
are produced with cross sections
of tens of femtobarns, for $1/R\simeq 500~$GeV. The
search for dijet resonances is a great challenge due to 
large backgrounds \cite{Abazov:2003tj, Simmons:1996fz},
but an observation at invariant masses consistent with 
the $t\overline{t}$ peaks 
would provide a further  confirmation of the models with
universal extra dimensions.

Here we have concentrated on single production of (1,1) modes.
The pair production of (1,0) modes is also interesting, and needs 
to be analyzed in detail. In the case of one universal extra dimension,
a search in the leptons plus missing-energy channel in Run I 
is already setting a limit of $1/R > 280$ GeV at the 95\% confidence level 
\cite{cdf-ued}. 
In order to set limits on two universal extra dimensions
based on pair production of
(1,0) modes, one needs to use the KK-number preserving interactions
derived in Ref.~\cite{Burdman:2005sr} 
and compute the relevant cross sections and branching fractions.
Compared to the case of one universal extra dimension,
the presence of spinless adjoints could substantially 
change both the cross sections for pair production 
\cite{Macesanu:2002db} and the branching fractions \cite{Cheng:2002ab}.
We leave this important study for future work.

\section{Prospects for the LHC}\label{sec:LHC}
\setcounter{equation}{0}

By contrast to the Tevatron, where the dominant contribution to
the production of $(1,1)$ modes comes from $q\overline{q}$ 
annihilation, at the LHC there will be competing contributions
from parton-level processes involving gluons in the initial state.
The KK-number violating couplings of gluons arise from 
higher-dimensional operators generated at the one-loop level. 
Although these couplings are not enhanced by a logarithmic factor, as 
the $q\overline{q}$ couplings to vector modes
(see section \ref{sec:KKcouplings}), the presence of the 
gluon in the initial state may compensate this effect due to a larger 
PDF at moderate energies.
The main processes are 
$s$-channel production due to dimension-6 operators of the   
$G^{(1,1)}_\mu$ and $B_\mu^{(1,1)}$
vector modes through gluon fusion,
and of the (1,1) quark modes through quark-gluon fusion.
It is beyond the scope of this article to compute the coefficients
of these operators which arise from finite one-loop contributions. 
In order to have an order of magnitude estimate
of (1,1)-mode production at the LHC 
we compute the $q\overline{q}$ annihilation processes
which have logarithmic enhancements of the couplings 
but smaller PDF's for $\overline{q}$.

\begin{figure}[t]
\hspace*{-.5em}
\psfrag{sigma}[b]{\hspace*{1.5em} $\sigma 
\left(\rule{0em}{0.6em}\right. pp\rightarrow 
V^{(1,1)}_\mu \!\left.\rule{0em}{0.6em}\right)$ \ [fb]}
\psfrag{MR}[t]{\hspace*{.9em} $M_{V^{(1,1)}}$ [TeV]}
\psfig{file=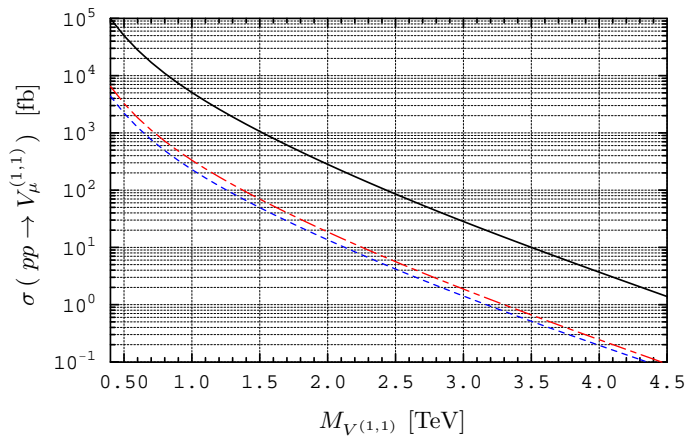,width=8.9cm,angle=0}
\vskip0.1cm
\caption{Production cross sections for $(1,1)$ vector modes in the 
$s$ channel at the LHC due to $q\overline{q}$ annihilation.
The solid,  dashed and dotted (lowest) lines
represent the $G^{(1,1)}_\mu$,  $W^{(1,1)3}_\mu$ and  $B^{(1,1)}_\mu$ 
production cross sections, respectively.} 
\label{fig:LHC}
\end{figure}

The production cross sections at the LHC for $G_\mu^{(1,1)}$,
$W_\mu^{(1,1)3}$, and $B_\mu^{(1,1)}$ due to $q\overline{q}$
annihilation are given by the right-hand sides
of Eqs.~(\ref{g11xs}) and (\ref{b11xs}) with $t_q(z)$ replaced by
\be
\int_{z}^1 \frac{dx}{x}\left[ \rule{0mm}{3.5mm}  
q(x)\,\overline{q}\left(z/x\right) + \overline{q}(x)\, q\left(z/x\right) 
\rule{0mm}{3.5mm}  \right] ~,
\ee
to leading order in $\alpha_s$. 
In Figure 8 we plot these three cross sections, using the CTEQ6 
PDF's at leading order with a $K$ factor of 1.3.
In the case of $G_\mu^{(1,1)}$,
and $B_\mu^{(1,1)}$, these are underestimates of the 
total production cross sections because  
the additional contributions
to the production cross sections from gluon fusion mentioned above
are not included.
For $W_\mu^{(1,1)3}$ production, the $SU(2)_W$ gauge symmetry 
does not allow gluon fusion via dimension-6 operators, and therefore the
only relevant parton-level process is $q\overline{q} \to W_\mu^{(1,1)3}$.
One should keep in mind though that 
additional $W_\mu^{(1,1)3}$ particles are produced from the cascade decays
of $G_\mu^{(1,1)}$, as explained in Section \ref{sec:GDecays}.

In order to translate these high rates at the LHC into a mass reach 
it is necessary to study carefully the backgrounds, which are 
huge for values of $1/R$ in the few hundred GeV, where
the Tevatron has a significant discovery potential. 
For larger values of the
compactification scale, the backgrounds should be manageable
for the $t \bar t$ signal.  
Moreover, for large $1/R$ the
decay of $G^{(1,1)}_\mu$ to a top quark and its (1,1) mode
opens up (see Section \ref{sec:GDecays}), 
leading to additional interesting signals 
involving $t$ and $b$ quarks.
Thus, the LHC will complement the searches at the Tevatron
discussed in Section \ref{sec:tevatron}, 
by probing larger values of $1/R$.

In Figure~9 we plot the cross sections for $t\bar t$ pairs
coming from the $G_H^{(1,1)}+W_\mu^{(1,1)3}$,
$W_H^{(1,1)3}+B_\mu^{(1,1)}$, and $B_H^{(1,1)}$ 
mass peaks, including only the $q\overline{q}$ initial states. 
Comparing these cross sections to the discovery potential of the ATLAS
detector for a narrow resonance decaying to $t\overline{t}$,
given in \cite{unknown:1999fr, Beneke:2000hk},
we estimate that the production cross sections
for (1,1) modes are large enough 
to allow discovery of narrow $t\overline{t}$ resonances 
of at least 1 TeV with an integrated luminosity of 30 fb$^{-1}$.
The reach can be further increased by using the extra leptons 
produced in the cascade decays of the $W_\mu^{(1,1)3}$ and  $W_\mu^{(1,1)\pm}$ 
modes, as shown in Figure 5.

If a discovery is made, further measurements may be performed:
angular distributions, threshold effects in cascade decays, 
lepton pairs from $B^{(1,1)}_\mu$ decays (the branching fraction
is 1\% for each of the $e^+e^-$ and $\mu^+\mu^-$ pairs).
A thorough study of the capabilities of the LHC, both in the
hadronic and leptonic channels is needed.  
Particularly exciting would be to identify the spinless 
adjoints, since the presence of these states is a distinctive 
feature of the 6D scenario.

\begin{figure}
\psfrag{sigma}[B]{\hspace*{2em} $\sigma
\left(\rule{0em}{0.5em}\right. pp\rightarrow V^{(1,1)}
 \rightarrow t\overline{t} \left.\rule{0em}{0.5em}\right) $ \ [fb]}
\psfrag{Mtt}[t]{$M\!\left(t\overline{t}\right)$ \ [TeV]}
\hspace*{-0.9em}
\psfig{file=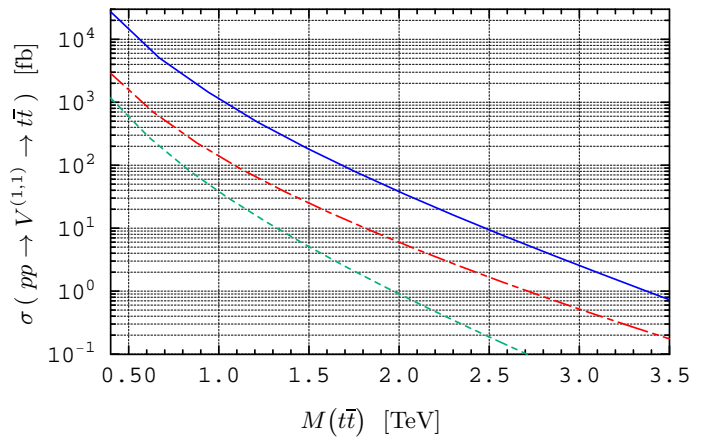,width=9.1cm,angle=0}
\vspace*{-1mm}
\caption{Cross section for the production of $t\bar t$ pairs
at the LHC from the $G_H^{(1,1)}+W_\mu^{(1,1)3}$ (top, solid line),
$W_H^{(1,1)3}+B_\mu^{(1,1)}$ (middle line), and $B_H^{(1,1)}$
(bottom line) peaks.}
\label{x112q-LHC}
\end{figure}

The most convincing test of the existence of
two universal extra dimensions would be the observation of 
series of resonances clustered around the masses of the
$(j,k)$ levels with $j+k$ even. 
Relative to the first even level, of mass
$M_{1,1} = \sqrt{2}/R$, the next four even levels have masses
$M_{2,0} = \sqrt{2} M_{1,1}$,
$M_{2,2} = 2 M_{1,1}$,
$M_{3,1} = \sqrt{5} M_{1,1}$, 
$M_{4,0} = 2\sqrt{2} M_{1,1}$.
Within each of these levels, the relative mass splittings 
are roughly the same as for the (1,1) level (see Figure~\ref{fig:second}).
However, the branching fractions into zero-mode fermions are smaller 
than for the corresponding 
(1,1) mode because the higher level KK modes may also 
decay into lower level ones.

\bigskip

\section{Conclusions}\label{sec:conclusions} 
\setcounter{equation}{0}

The 6D Standard Model compactified on the chiral square 
\cite{Burdman:2005sr,Dobrescu:2004zi} is a well motivated theory,
given that it predicts a long proton lifetime \cite{Appelquist:2001mj},
it restricts the number of fermion generations to a multiple of three
\cite{Dobrescu:2001ae}, and it accommodates nicely the observed pattern
of neutrino oscillations \cite{Appelquist:2002ft}.
We have computed here the spectrum of KK modes,  which is
split due to localized operators induced by one-loop effects
(see Section \ref{sec:6DSM}). In particular, we have shown that the 
lightest KK particle in this model is the hypercharge spinless 
adjoint $B_H^{(1,0)}$, whose mass is roughly 15\% below
the compactification scale $1/R$ which sets the tree-level mass
of the (1,0) KK modes.
This appears to be a promising dark matter candidate,
but in order to find the range of values for $1/R$
consistent with the dark matter abundance one would need to 
determine the relic density along the lines of 
the detailed computations performed in the case of one universal 
extra dimension \cite{Kong:2005hn,Kakizaki:2005uy}.

We have also computed the KK-number violating interactions due to 
loop-induced localized operators which, 
although suppressed compared to the tree level interactions presented 
in Ref.~\cite{Burdman:2005sr}, have important phenomenological 
consequences.
In this way we have laid the groundwork for studies  
of the phenomenology of two universal extra dimensions.

After completing this general study of the KK couplings and masses,
we have focused on the (1,1) modes, which are even under 
KK-parity, and therefore may be produced in the $s$ channel at 
colliders. The (1,1) modes are the lightest KK modes of this type,
with a tree level mass of only $\sqrt{2}/R$.
This low mass for an even KK level, and the presence of spinless adjoints 
changes significantly the phenomenology 
compared to the case of level-2 modes from one universal extra dimension 
discussed in Refs.~\cite{Cheng:2002ab, Datta:2005zs}.

In Section \ref{sec:decays} we have computed the branching 
fractions of the $(1,1)$ KK modes.
As in the case of one universal extra dimension, the even 
vector modes are leptophobic because the loop-induced couplings
to zero-mode  leptons are generated by the $SU(2)_W\times U(1)_Y$
interactions, while the loop-induced couplings
to zero-mode quarks are generated by the $SU(3)_C$
interactions. Only the hypercharge (1,1) mode has a non-negligible
branching fractions into lepton pairs: 1\% for each of $e^+e^-$
and $\mu^+\mu^-$.
An interesting result of our computation is that the branching fractions
to $t\overline{t}$ pairs are enhanced, especially because the strength
of the (1,1) couplings to the top quark is increased by one-loop corrections 
involving the Yukawa interaction to the Higgs fields.
Even more strikingly, the spinless adjoints decay most of the time into
$t\overline{t}$ pairs, because their couplings to zero-mode
fermions are proportional to the fermion mass. Putting together the 
direct decays and cascade decays of vector (1,1) modes, we have found 
large branching fractions for final states involving $t\overline{t}$
resonances: 72\%, 21\% and 26\% for $G_\mu^{(1,1)}$,
$W_\mu^{(1,1)3}$ and $B_\mu^{(1,1)}$, respectively, for
$1/R\sim 500$ GeV.

Although leptophobic bosons are usually hard to observe at 
hadron colliders, due to large backgrounds, the sizable branching fractions
into $t\overline{t}$ offer promising prospects for searches at 
the Tevatron and the LHC.
We have shown  that the Tevatron is likely to set useful limits on $1/R$,
through $s$-channel production of 
the (1,1) gluon, $B^{(1,1)}_\mu$ and $W^{(1,1)3}_\mu$, and their
subsequent cascade or direct decays to a pair of top quarks
(see Section \ref{sec:limit}).  
Altogether there are five narrow resonances to be 
observed in the invariant $t\overline{t}$ mass distribution, but 
they form at most three mass peaks once we take into 
account a realistic $t\overline{t}$  pair mass resolution.
With $4{\rm fb}^{-1}$, the D0 and CDF experiments may 
discover resonances in the 
$t\overline{t}$ channel, or else will likely set a lower limit on 
$t\overline{t}$ mass peaks in the 500--700 GeV range, corresponding 
to a lower limit on $1/R$ in the 320--450 GeV range. 

If a discovery of one or more $t\overline{t}$ resonances is made at the 
Tevatron, or for larger $1/R$ at the LHC, there are various other measurements
that can differentiate the 6D Standard Model from other theories,
such as Topcolor \cite{Hill:1991at}, that predict similar resonances.
Particularly useful would be measurements of the
extra jets and leptons from cascade decays,
angular distributions in the decays of spinless adjoints,
the dijet invariant mass distribution that may reveal 
resonances with the same mass as in the $t\overline{t}$ channel,
and signals involving missing transverse energy from pair 
production of (1,0) modes.

Despite the troublesome backgrounds at the LHC, the large rates for 
producing $t\overline{t}$ resonances at high invariant mass,
in the TeV range, would allow accurate measurements. For a precise assessment
of the LHC reach in $1/R$, more detailed studies are needed. 
Particularly exciting would be the discovery of resonances associated with 
several KK levels. The masses of consecutive even levels have ratios given 
by a peculiar 
factor of $\sqrt{2}$ for the first three even levels, so that the observation 
of clusters of resonances fitting this pattern would 
signal the existence of two universal extra dimensions.

\bigskip

{\bf Acknowledgments:} \ We are grateful to Robert Harris,
Jacobo Konigsberg and Greg Landsberg for illuminating
explanations regarding the capabilities of CMS, CDF and D0.
We would like to thank Elizabeth Simmons and 
Hitoshi Murayama for comments
regarding KK-mode production, and Frank Petriello for a couple
of useful observations.
G.B. acknowledges the support of the State of S\~{a}o Paulo
Research Foundation (FAPESP), and the Brazilian  National Counsel
for Technological and Scientific Development (CNPq).
The work of B.D. was supported by DOE under contract DE-FG02-92ER-40704. 
E.P. was supported by DOE under contract DE-FG02-92ER-40699.


\end{document}